\documentclass[apj,twocolumn,twocolappendix,floatfix]{openjournal}

\DeclareRobustCommand{\VAN}[3]{#2}
\let\VANthebibliography\thebibliography
\def\thebibliography{\DeclareRobustCommand{\VAN}[3]{##3}\VANthebibliography}

\usepackage{amsmath}	% Advanced maths commands
\usepackage{amssymb}	% Extra maths symbols
\usepackage{balance}
\usepackage{booktabs}
\usepackage{lipsum}
\usepackage[T1]{fontenc}
\usepackage{graphicx}	% Including figure files
\usepackage[breaklinks,colorlinks,citecolor=blue,urlcolor=blue,linkcolor=blue]{hyperref}
\usepackage{cleveref}
\usepackage{multirow}
\usepackage{newtxtext,newtxmath}
\usepackage{orcidlink}
\usepackage{pifont}% http://ctan.org/pkg/pifont
\usepackage[nobottomtitles]{titlesec}
\usepackage{natbib}
\usepackage{xparse}
\usepackage{ifthen}
\usepackage{gensymb}
\usepackage{ctable}
\titleformat{\section}{\filcenter\MakeUppercase}{\thesection.}{0.5em}{}

\usepackage{savesym}
\savesymbol{tablenum}
\usepackage{siunitx}
\restoresymbol{SIX}{tablenum}
\usepackage{needspace}
\usepackage{overpic}
\usepackage{xcolor}
\usepackage{lipsum}
\usepackage{pifont}

\newcommand{\cmark}{\ding{51}}%
\newcommand{\xmark}{\ding{55}}%

\begin{document}
% Title of the paper, and the short title which is used in the headers.
% Keep the title short and informative.
\title[Astronomical Cardiology]{Astronomical Cardiology: A Search For Heartbeat Stars Using \textit{Gaia} and \textit{TESS}}

% The list of authors, and the short list which is used in the headers.
% If you need two or more lines of authors, add an extra line using \newauthor

\author{\vspace{-1.3cm}Jowen Callahan,$^{1}$,
D. M. Rowan\,\orcidlink{0000-0003-2431-981X}$^{1,2}$,
C. S. Kochanek\ $^{1,2}$,
        K. Z. Stanek\,\orcidlink{0009-0001-1470-8400}$^{1,2}$,
}

\affiliation{$^{1}$Department of Astronomy, The Ohio State University, 140 West 18th Avenue, Columbus, OH, 43210, USA\\
$^{2}$Center for Cosmology and Astroparticle Physics, The Ohio State University, 191 W. Woodruff Avenue, Columbus, OH, 43210, USA
}

% Abstract of the paper
\begin{abstract}
Heartbeat stars are a subclass of binary stars with short periods, high eccentricities, and phase-folded light curves that resemble an electrocardiogram. We start from the \textit{Gaia} catalogs of spectroscopic binaries and use \textit{TESS} photometry to identify 112 new heartbeat star systems. We fit their phase-folded light curves with an analytic model to measure their orbital periods, eccentricities, inclinations, and arguments of periastron. We then compare these orbital parameters to the \textit{Gaia} spectroscopic orbital solution. Our periods and eccentricities are consistent with the \textit{Gaia} solutions for 85$\%$ of the single-line spectroscopic binaries but only 20$\%$ of the double-line spectroscopic binaries. For the two double-line spectroscopic binary heartbeat stars with consistent orbits, we combine the \textit{TESS} phase-folded light curve and the \textit{Gaia} velocity semi-amplitudes to measure the stellar masses and radii with \texttt{PHOEBE}. In a statistical analysis of the heartbeat star population, we find that non-giant heartbeat stars have evolved off the main sequence and that the fraction of the systems that are heartbeat stars rises rapidly with effective temperature.
\end{abstract}

\maketitle

\section{Introduction}

Heartbeat stars (HB) are a relatively rare subclass of detached binary stars with short periods ($P\lesssim100$ days) and high eccentricities ($e\gtrsim0.2$). In these binaries, the stars are tidally deformed near pericenter, causing periodic brightness variations. HBs are easily identifiable because their phase-folded light curves resemble an electrocardiogram \citep{Thompson2012}, and a simple analytic model \citep{Kumar1995} can be used to estimate the properties of the orbit. The tides can also produce tidally excited oscillations (TEOs) of the star that are observed between pericenter passages.

Since the amplitude of the photometric variations in HBs is typically small, it was only with the \textit{Kepler} mission that HBs were detected \citep{Welsh2011}. KOI-54 was one of the first HBs from \textit{Kepler}, and it also had TEOs \citep{Welsh2011, Fuller2012, Burkart2012, Leary2014}. However, HBs would not have their name until \citet{Thompson2012} applied the analytic \citet{Kumar1995} model to fit the \textit{Kepler} phase-folded light curves of 17 new HB systems. Hundreds of HBs were identified in \textit{Kepler} \citep{Hambleton2013, Kirk2016, Shporer2016}. These systems are all on the main sequence (MS) and have periods of $10\lesssim P\lesssim100$ days. The Optical Gravitational Lensing Experiment (OGLE) has discovered nearly 1000 HBs across the upper MS and giant branch \citep{Wrona2022a, Wrona2022b}.

The \textit{Transiting Exoplanet Survey Satellite} \citep[\textit{TESS}, ][]{Ricker2015} provides another opportunity to discover short period ($P<10$ days) HBs. \textit{TESS} observes the sky in twenty-six sectors which are continuously observed for roughly twenty-seven days. Like in \textit{Kepler}, early detections of HBs in \textit{TESS} came as a byproduct of planetary searches \citep{Wheel2019}. \textit{TESS} photometry has also been used to identify or characterize HB variability in known binaries like MACHO 80.7443.1718 \citep{THARINDU2019, THARINDU2021, SK2022}, KIC 5006817 \citep{Merc2021}, V680 Mon \citep{Paunzen2021}, WR 21a \citep{Barb2022}, and FX UMa \citep{Wang2023}. HBs are also discovered serendipitously in various \textit{TESS} star samples \citep{Murphy2020, Kochukhov2021, Sharma2022, PMR2024}.

There have also been systematic surveys for HBs in \textit{TESS}. By analyzing the first dozen \textit{TESS} sectors, \citet{SK2021} found 20 new HB systems through the visual inspection of the phase-folded light curves of spectroscopic binaries. \citet{Li2024a, Li2024b} doubled the number of HBs found by deliberate searches in \textit{TESS} by cross-matching and visually inspecting the \textit{TESS} light curves of the stars in the PPM catalog \citep{1993Rser, 1994Rser, 1995Nest}. Recently, \citet{Sol2025} identified 180 HBs using the \textit{TESS} full-frame images by applying convolutional neural networks to TESS phase-folded light curves, followed by a visual inspection to confirm the candidates.

In contrast to previous HB searches, which start from large samples of phase-folded light curves, here we start from the \textit{Gaia} catalogs of double-line (SB2) and single-line (SB1) spectroscopic binaries. This has the advantage of starting from a sample of known binaries, and we can then use the SB2 HB systems to estimate the masses and radii of the stars. In Section~\S\ref{sec:HBS}, we discuss the \textit{Gaia} data and how we model and search their phase-folded \textit{TESS} light curves for HBs. Sections~\S\ref{sec:SB2S} and~\S\ref{sec:SB1S} describe our search strategies for HBs in the \textit{Gaia} SB2 and SB1 samples, respectively. In Section~\S\ref{sec:GC}, we compare the orbital parameters we measure from the \textit{TESS} phase-folded light curves to the \textit{Gaia} orbital solutions. We also model the SB2 HBs with consistent \textit{Gaia}/HB orbital parameters to determine the masses and radii of the systems. Finally, Section~\S\ref{sec:D} discusses the statistics of HBs and future steps.

\begin{figure*}
	\includegraphics[width=\textwidth]{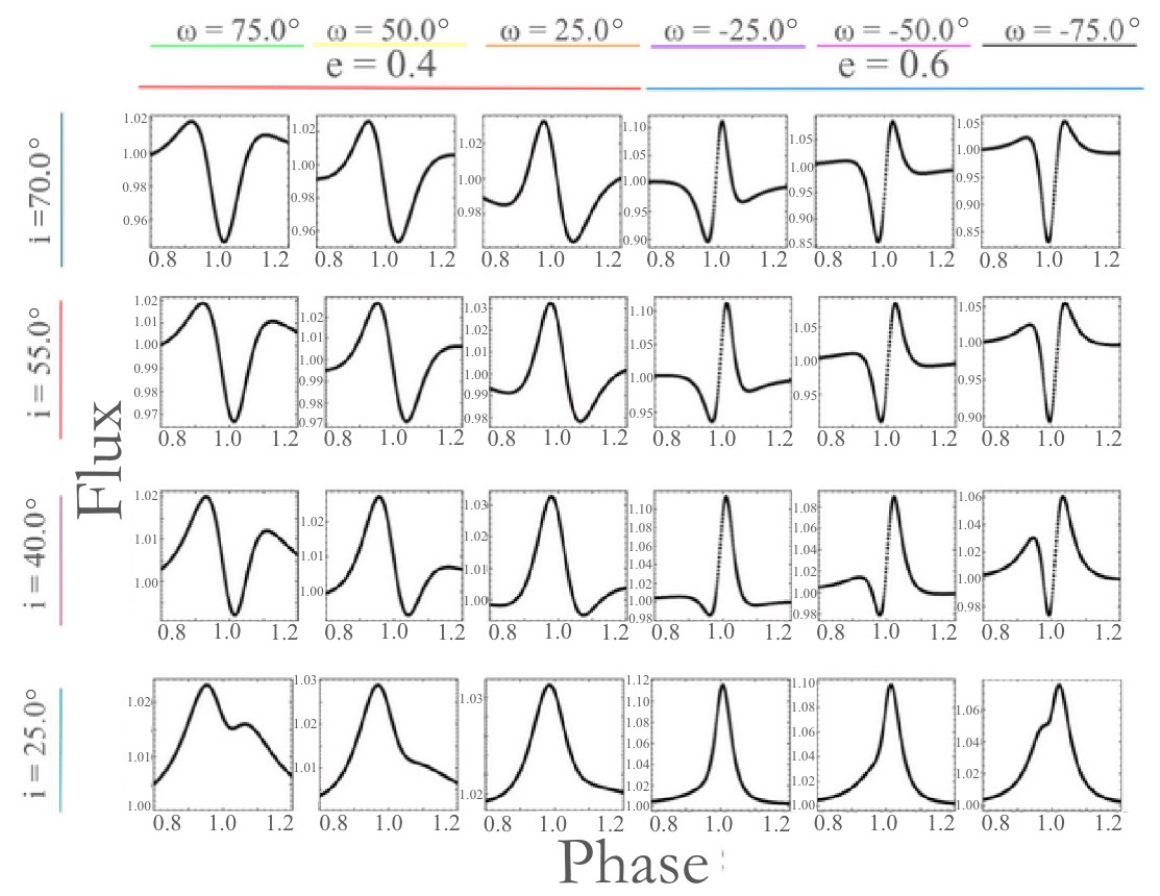}
    \caption{Examples of heartbeat star phase-folded light curves created using the \citet{Kumar1995} model, varying the orbital eccentricities (e$=0.4$, left three columns and e$=0.6$, right three columns), for a range of arguments of periastron (horizontal) and inclinations (vertical). This Figure is based on Figure 5 of \citet{Thompson2012}.}
    \label{fig:Tharindu}
\end{figure*}

\section{Heartbeat Search}
\label{sec:HBS}

\begin{figure*}
    \centering
    \begin{tabular}{ccc}
    \vspace{-1em}
	\includegraphics[width=0.33\textwidth]{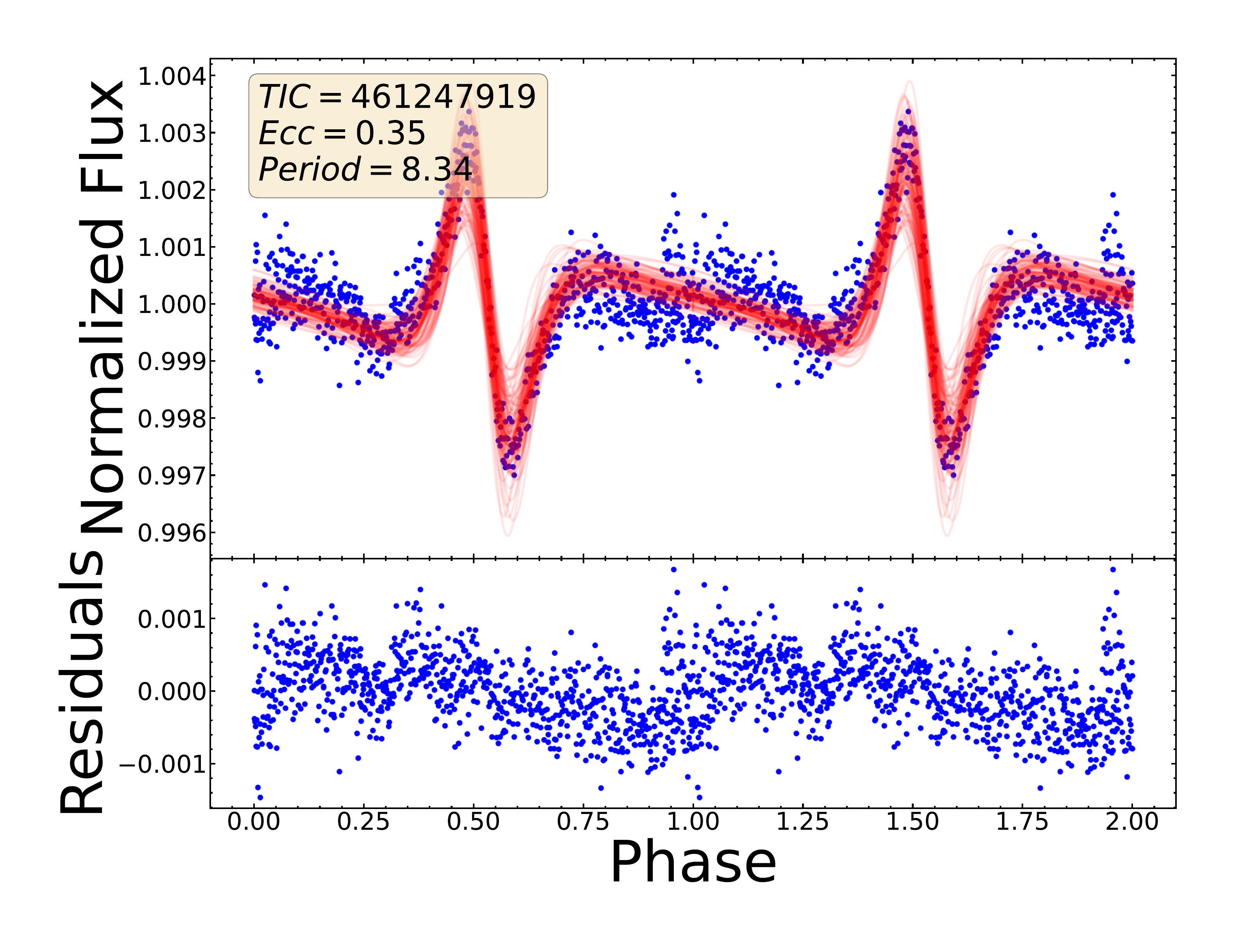} & 
    \hspace{-2em}
       \includegraphics[width=0.33\textwidth]{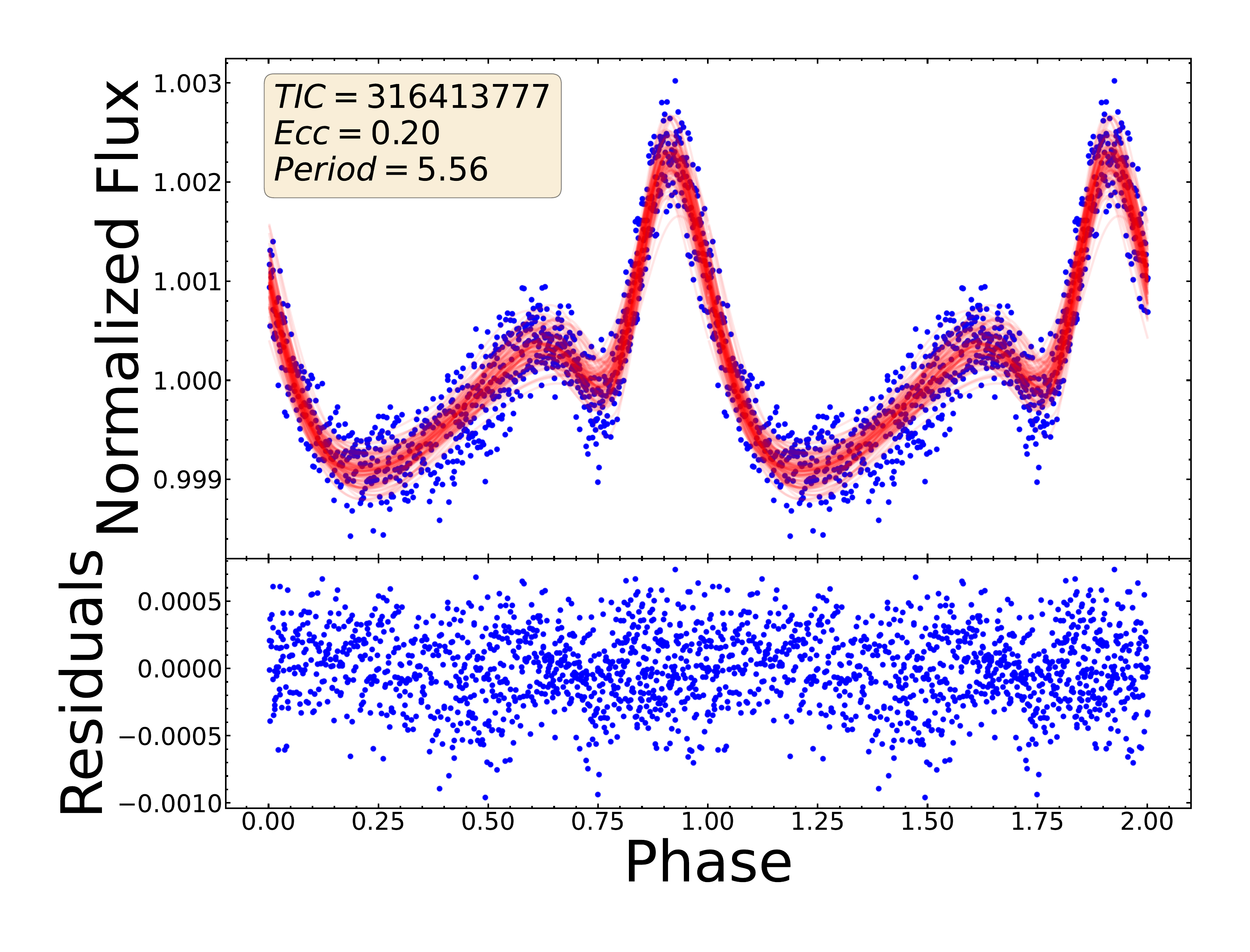} &
       \hspace{-2em}
        \includegraphics[width=0.33\textwidth]{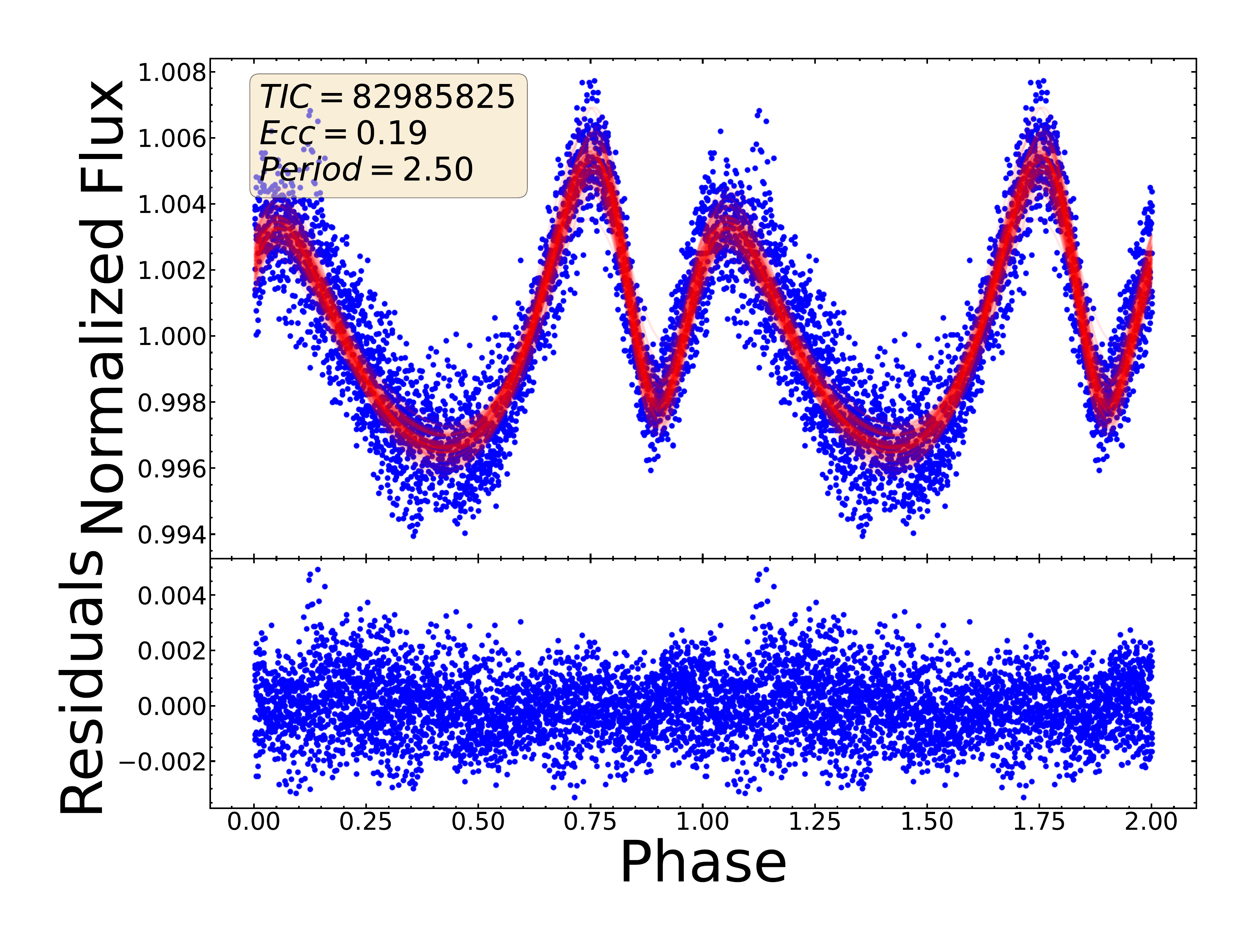} \\
        \vspace{-1em}
        \includegraphics[width=0.33\textwidth]{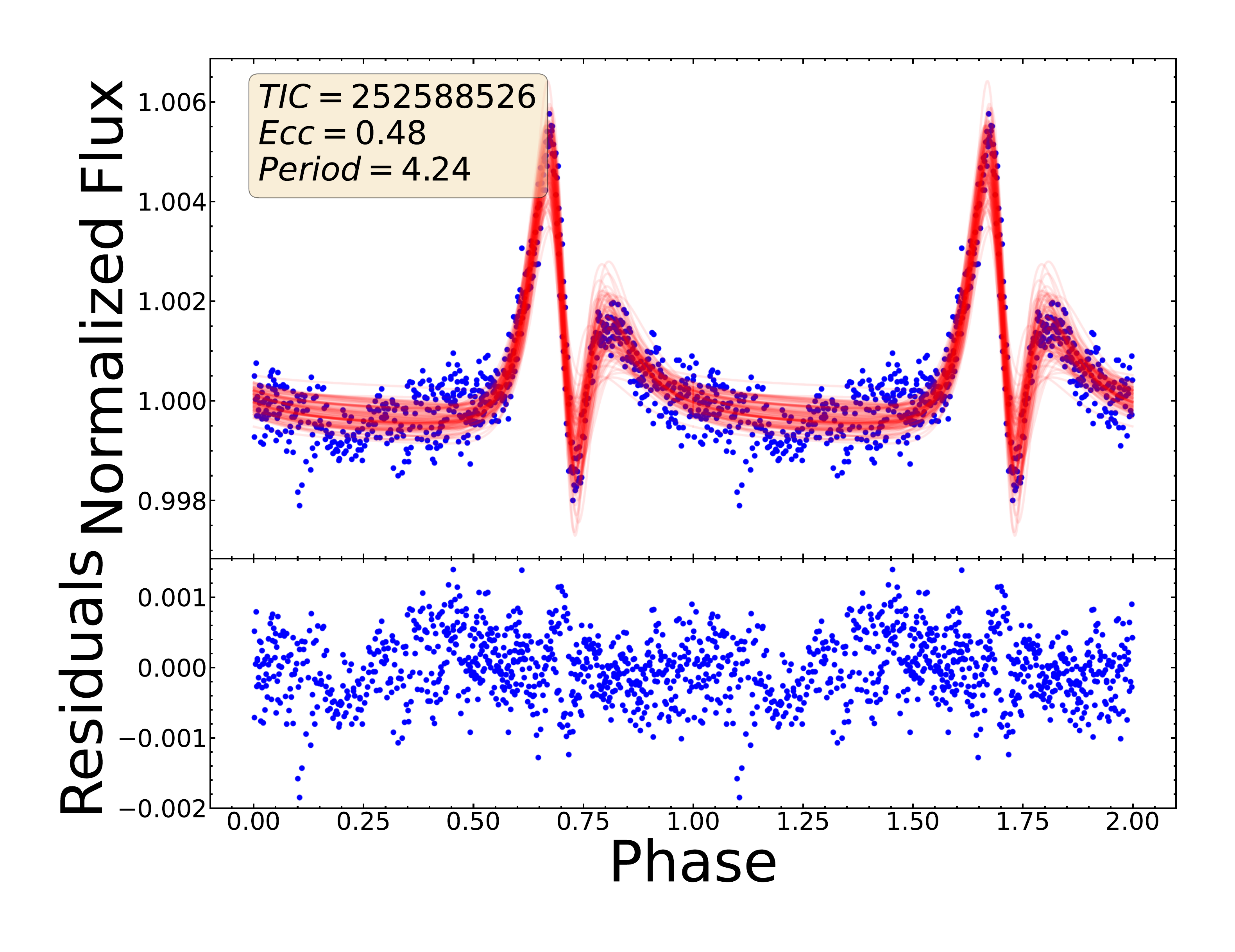} &
        \hspace{-2em}
        \includegraphics[width=0.33\textwidth]{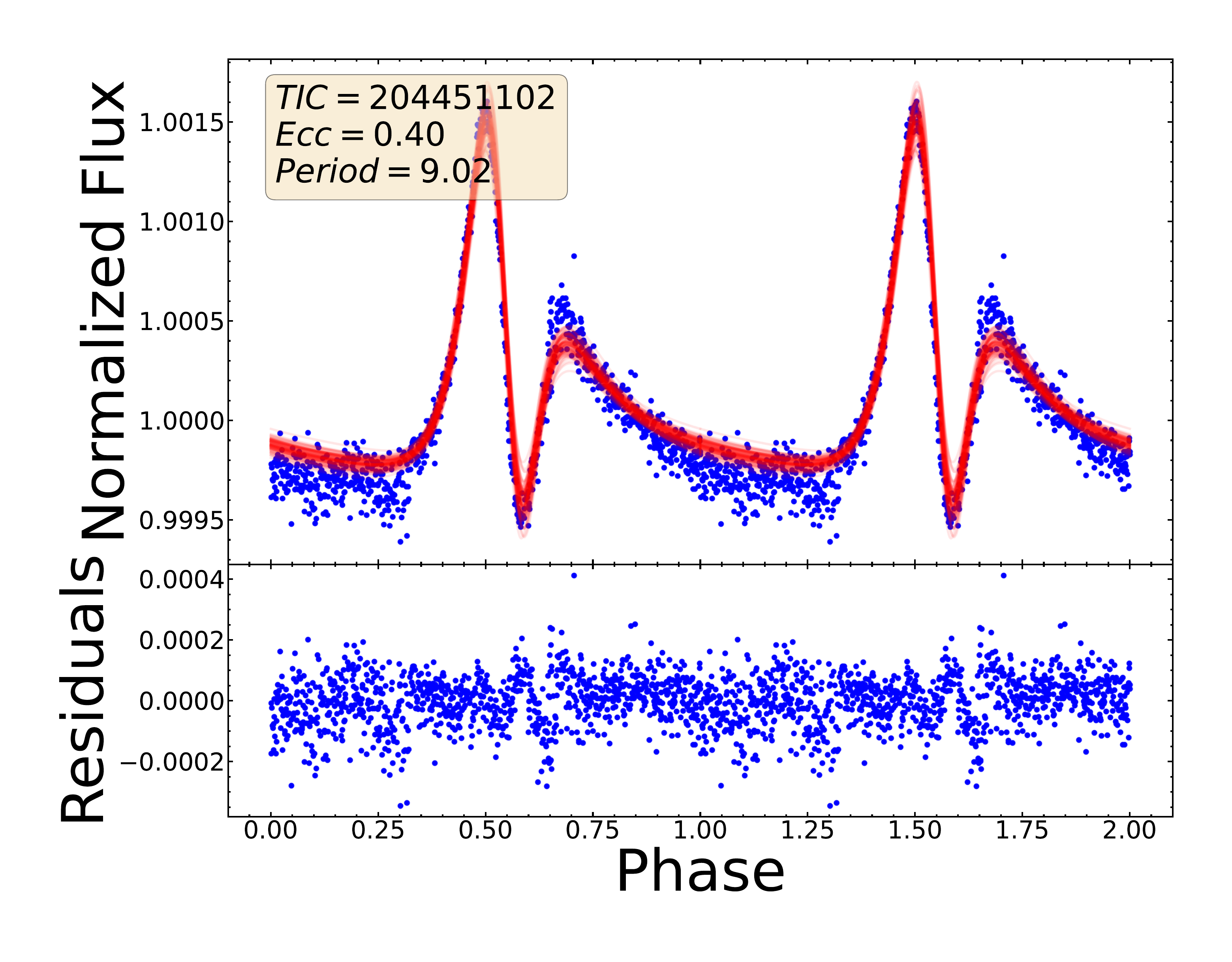} &
        \hspace{-2em}
        \includegraphics[width=0.33\textwidth]{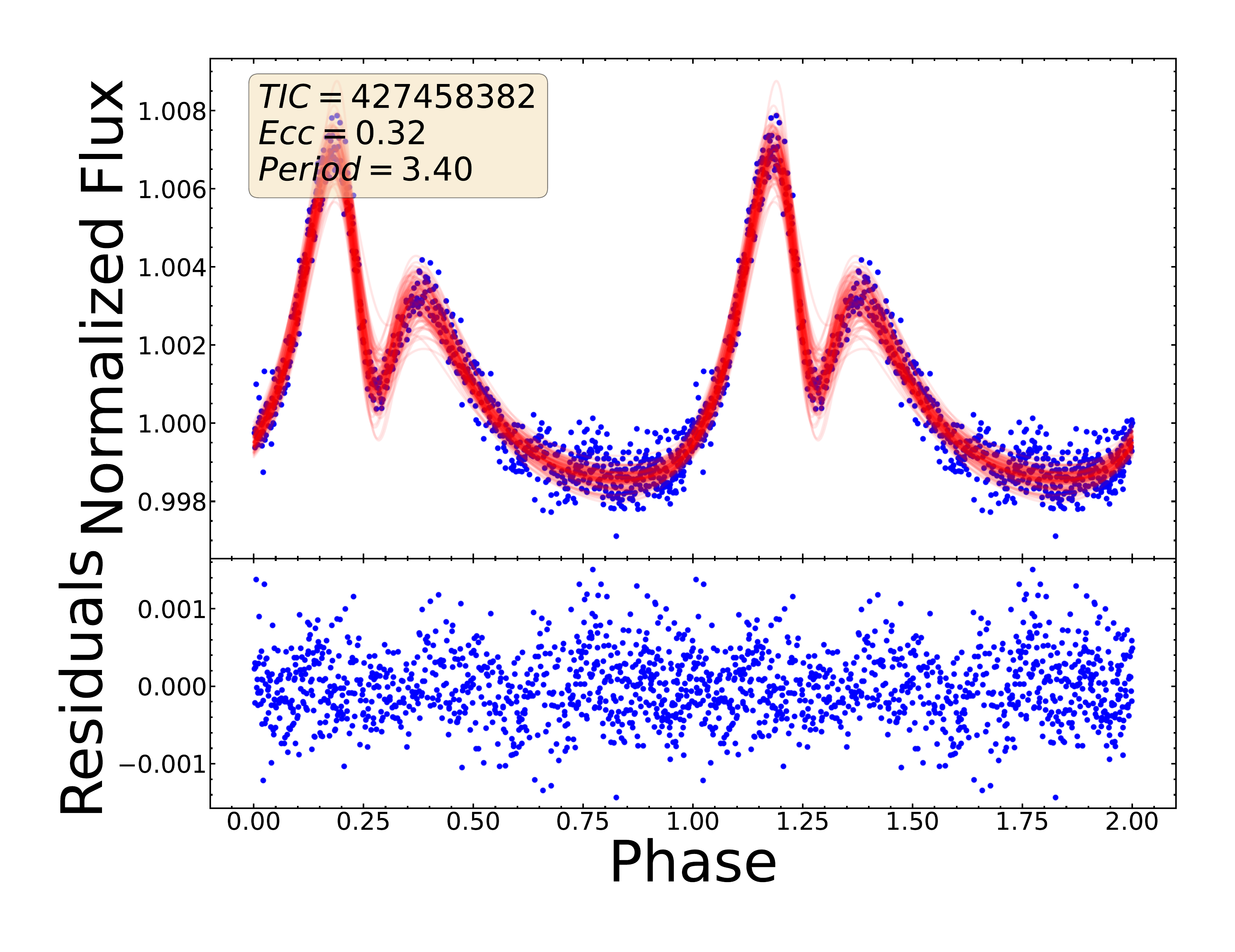} \\
        \vspace{-1em}
        \includegraphics[width=0.33\textwidth]{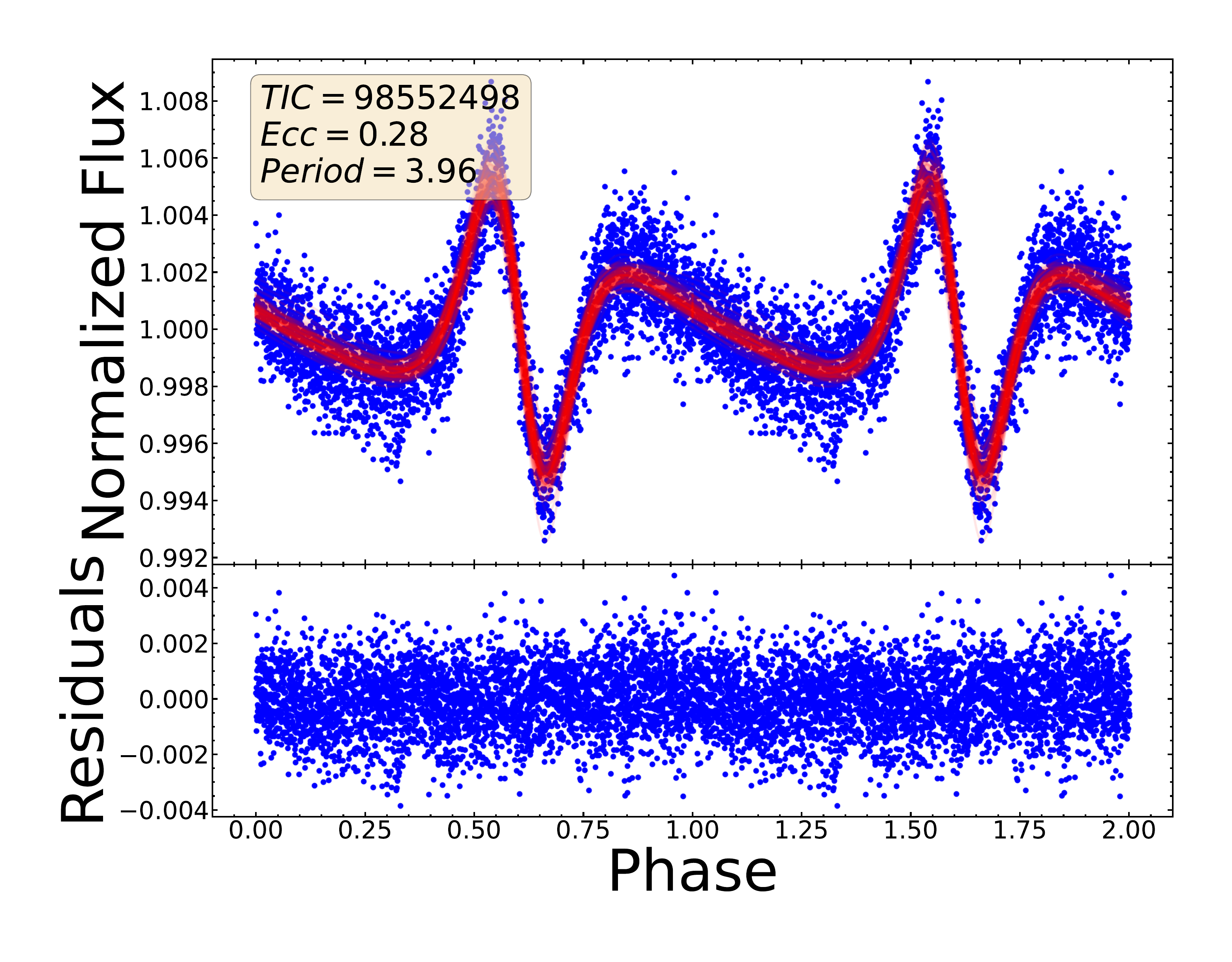} &
        \hspace{-2em}
        \includegraphics[width=0.33\textwidth]{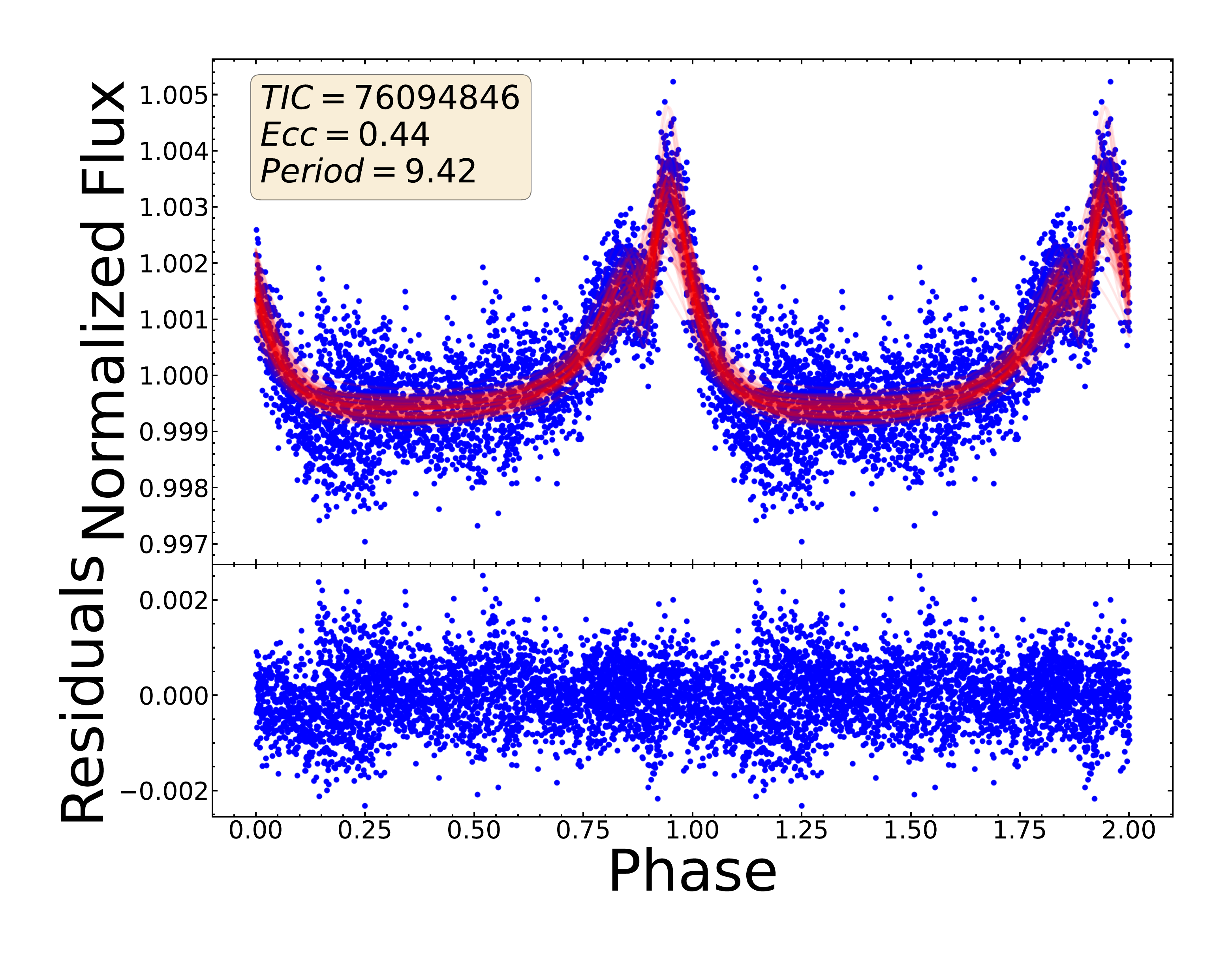} &
        \hspace{-2em}
        \includegraphics[width=0.33\textwidth]{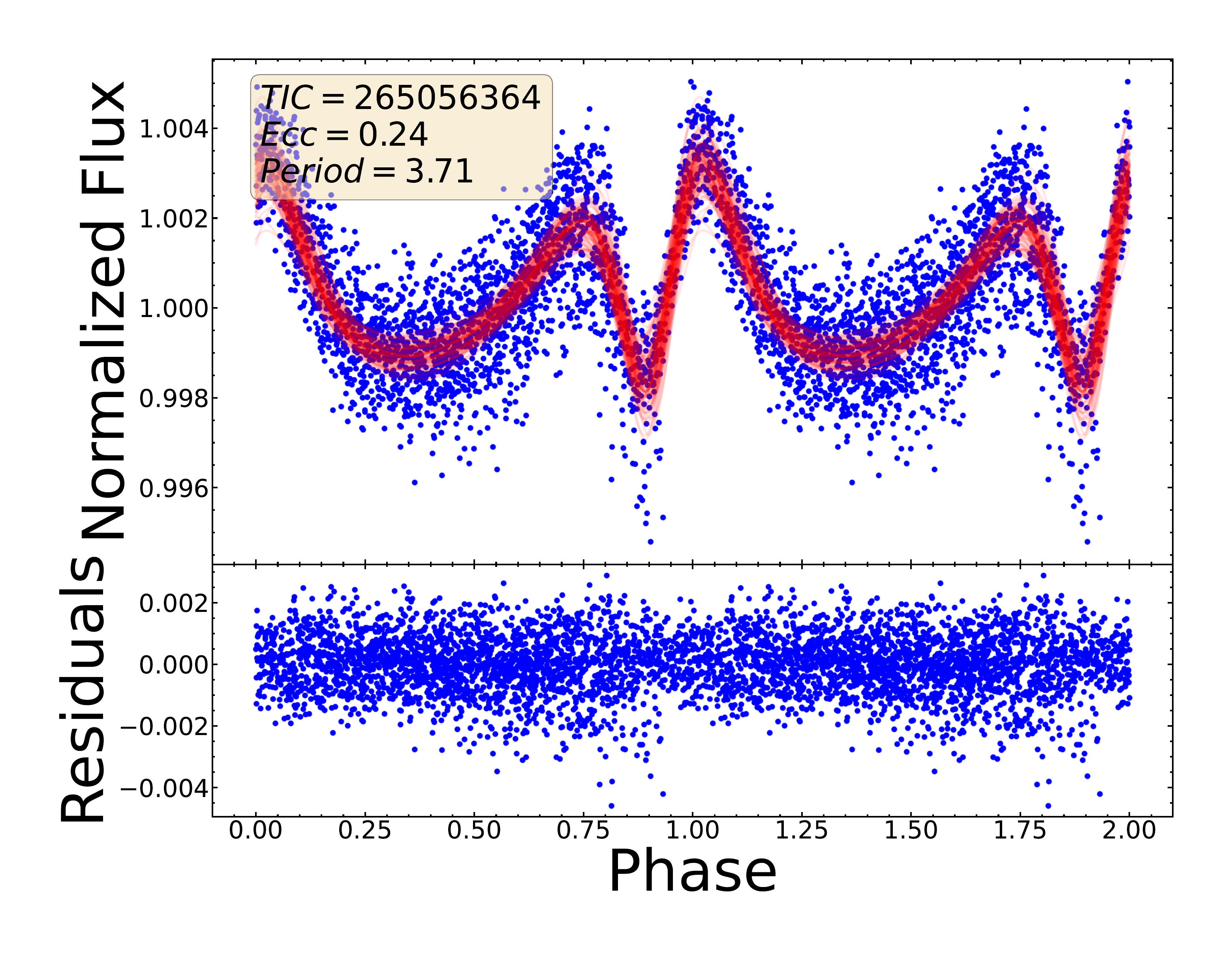} \\
    \end{tabular}
    \caption{The phase-folded light curves (upper panel) and the model residuals (lower panel) for nine of the double-line spectroscopic heartbeat binaries. The lines are the \citet{Kumar1995} models. The text boxes show the \textit{TESS} input catalog (TIC), eccentricity, and period (days) of each target.}
    \label{fig:LC}
\end{figure*}

We start from the 181,529 SB1s and 5,376 SB2s \citep{Gaia2023SB} in \textit{Gaia} DR3. We used phase-folded light curves for sectors 1-79 from the \textit{TESS} Quick-Look Pipeline \citep[QLP, ][]{Huang2020} for all of the \textit{Gaia} SB1s and SB2s. The typical \textit{Gaia} spectroscopic binary has four \textit{TESS} sectors of observations.

\subsection{SB2 Modeling}
\label{sec:SB2S}

\begin{table*}
	\centering
	\caption{The results of the light curve curve fit, \textit{Gaia} score, period, and eccentricity, flags for whether the orbital parameters agree, the existence of eclipses, and TEOs with the orbital harmonic ($n$). The full table is available in the file SBTABLE.full.}
    \begin{tabular}{| r  r  r  r  r  r  r  r  r  r  r  c  c  c  c |}
        \hline
		TIC & Type & Score & incl ($\degree$) & $\omega$ ($\degree$) & P (days) & \textit{Gaia} P (days) & P frac diff & e & \textit{Gaia} e & \textit{Gaia} e err & Match & Eclipses & TEOs & $n$\\	
        \hline

1598625 & SB1 & 0.442 & 36.130 & 232.130 & 3.252 & 3.241 & 0.004 & 0.330 & 0.404 & 0.096 & \cmark & \xmark & \xmark & $-$\\
9175061 & SB1 & 0.476 & 55.720 & 141.410 & 2.900 & 2.900 & 0.000 & 0.160 & 0.147 & 0.034 & \cmark & \xmark & \cmark & 2\\
11578775 & SB1 & 0.236 & 48.110 & 115.970 & 5.327 & 5.331 & 0.001 & 0.180 & 0.180 & 0.104 & \cmark & \xmark & \xmark & $-$\\
13933268 & SB1 & 0.374 & 51.480 & 308.690 & 4.129 & 4.126 & 0.001 & 0.110 & 0.127 & 0.019 & \cmark & \xmark & \xmark & $-$\\
15257961 & SB1 & 0.468 & 38.260 & 360.000 & 7.748 & 7.763 & 0.002 & 0.110 & 0.137 & 0.053 & \cmark & \xmark & \xmark & $-$\\
22567490 & SB1 & 0.320 & 56.800 & 61.110 & 6.567 & 6.554 & 0.002 & 0.550 & 0.250 & 0.034 & \xmark & \cmark & \xmark & $-$\\
26412885 & SB1 & 0.474 & 75.900 &  85.750 & 3.490 & 3.490 & 0.000 & 0.160 & 0.108 & 0.045 & \cmark & \xmark & \xmark & $-$\\
26689977 & SB1 & 0.262 & 40.260 & 301.160 & 3.692 & 3.697 & 0.001 & 0.110 & 0.125 & 0.053 & \cmark & \xmark & \xmark & $-$\\
28543727 & SB1 & 0.562 & 90.000 &  26.480 & 2.576 & 2.579 & 0.001 & 0.190 & 0.130 & 0.022 & \xmark & \xmark & \xmark & $-$\\
29451470 & SB1 & 0.497 & 65.910 & 343.170 & 3.011 & 3.260 & 0.076 & 0.110 & 0.102 & 0.043 & \cmark & \xmark & \cmark & 2\\
29518898 & SB1 & 0.376 & 90.000 & 306.270 & 4.649 & 4.628 & 0.004 & 0.130 & 0.163 & 0.074 & \cmark & \xmark & \xmark & $-$\\
31096993 & SB1 & 0.219 & 33.590 & 356.980 & 6.019 & 6.018 & 0.000 & 0.160 & 0.191 & 0.031 & \cmark & \xmark & \xmark & $-$\\
42430804 & SB1 & 0.239 & 45.180 & 216.590 & 8.998 & 0.403 & 21.314 & 0.560 & 0.168 & 0.060 & \xmark & \xmark & \xmark & $-$\\
42741990 & SB1 & 0.336 & 37.350 & 319.100 & 5.785 & 5.817 & 0.005 & 0.150 & 0.094 & 0.102 & \cmark & \xmark & \xmark & $-$\\
44787510 & SB1 & 0.720 & 22.360 & 360.000 & 3.553 & 1.860 & 0.910 & 0.470 & 0.096 & 0.019 & \xmark & \xmark & \xmark & $-$\\
61160125 & SB1 & 0.138 & 48.110 & 345.120 & 5.960 & 5.908 & 0.009 & 0.220 & 0.396 & 0.100 & \xmark & \xmark & \xmark & $-$\\
63548665 & SB1 & 0.286 & 50.370 & 298.410 & 4.770 & 1.124 & 3.243 & 0.190 & 0.057 & 0.047 & \xmark & \xmark & \xmark & $-$\\
65315466 & SB1 & 0.554 & 90.000 & 226.240 & 3.196 & 3.196 & 0.000 & 0.190 & 0.232 & 0.086 & \cmark & \xmark & \xmark & $-$\\
73486038 & SB1 & 0.771 & 57.060 &  85.460 & 6.445 & 6.451 & 0.001 & 0.260 & 0.302 & 0.048 & \cmark & \xmark & \xmark & $-$\\

\hline
\end{tabular}
	\label{tab:table}
\end{table*}

\begin{figure}
	\includegraphics[width=\columnwidth]{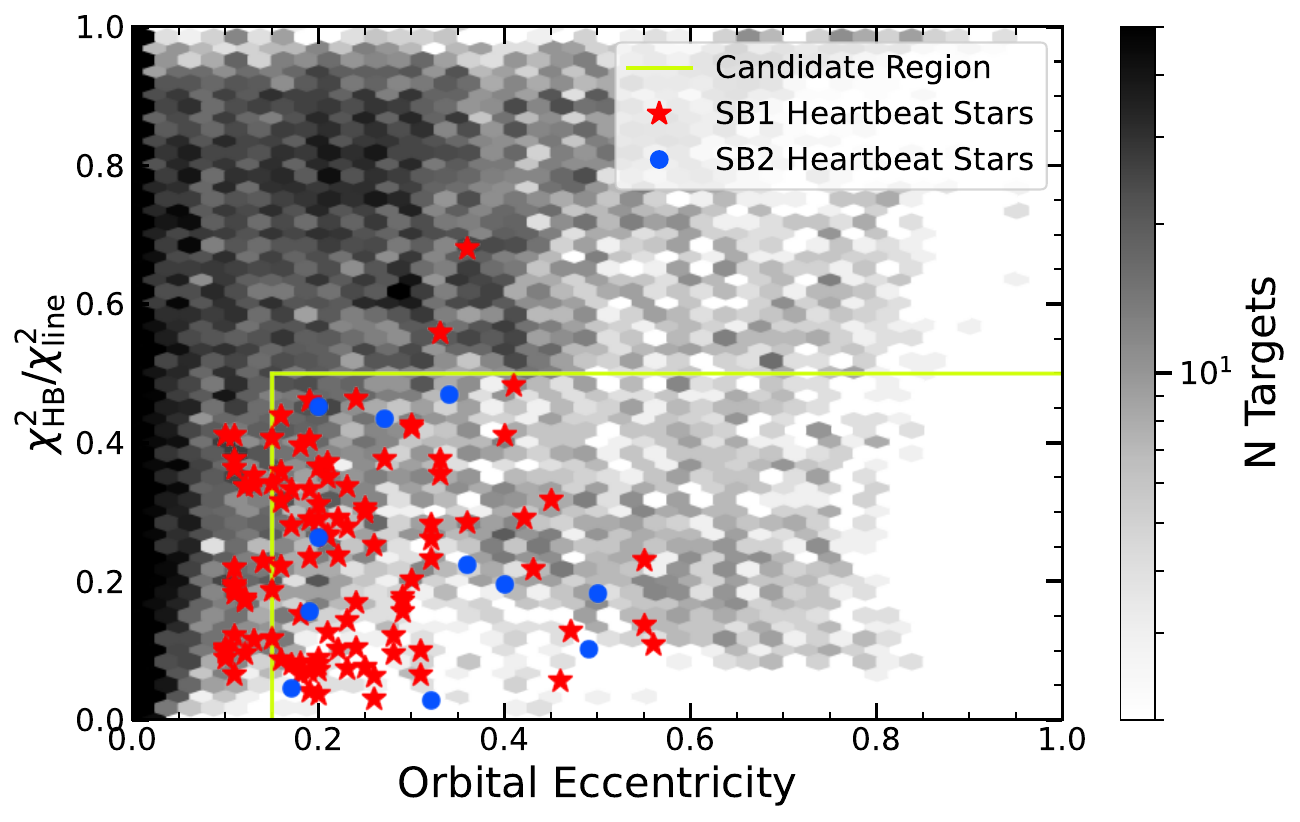}
    \caption{The density distribution of the $\chi^2$ ratio $R$ and eccentricity for the fits to all the \textit{Gaia} stars (grey background). We defined the selection region based on the results for the SB2 HBs (dots). The final models of the SB1 HBs (stars) can lie outside the selection region when masking eclipses leads to changes in the model.}
    \label{fig:Final}
\end{figure}

Since there are only 5,376 \textit{Gaia} SB2 targets, we identified HB candidates through visual inspection of the \textit{TESS} phase-folded light curves. For all targets, we run a Lomb-Scargle periodogram \citep[LS, ][]{Lomb1976, Scargle1982} on each \textit{TESS} sector independently. We visually inspected the phase-folded light curves and identified 10 HB candidates. We then use phase dispersion minimization \citep[PDM, ][]{Stellingwerf1978} to refine the orbital period. Before modeling the phase-folded light curve, we select a single \textit{TESS} sector for each target and remove long-term trends in the data, if necessary. For example, TIC 265056364 has a continuous decrease in flux over the course of the \textit{TESS} sector, which could be due to variable blended light or systematic trends. We remove these signals with a linear fit.

We fit the phase-folded light curves using the \citet{Kumar1995} analytic model. The flux is modeled as

\begin{equation}
    F=Z+S\frac{1-3\sin^2{i}\sin^2{(\nu+\omega)}}{(1-e\cos{E})^3},
	\label{eq:Flux}
\end{equation}where $S$ sets the amplitude, $Z$ is the mean flux, $\omega$ is the argument of periastron, $i$ is the inclination, and $e$ is the orbital eccentricity. The true anomaly is

\begin{equation}
    \nu={\frac{2\tan{(E/2)}}{\tan^{-1}{(\sqrt{1+e}/\sqrt{1-e})}}},
	\label{eq:True}
\end{equation}and the eccentric anomaly

\begin{equation}
    E-e\sin{E} = \frac{2\pi(t-t_{0})}{P}.
	\label{eq:Ecc}
\end{equation}is determined by the period, $P$ and the epoch of periastron, $t_0$. The shape of the phase-folded light curve varies significantly with the orbital configuration, as shown in Fig.~\ref{fig:Tharindu}.

We used Markov Chain Monte Carlo (MCMC) methods as implemented by \citet{FM2013} to sample over the parameters in the model. We ran the MCMC models with fourteen walkers, ten thousand iterations, and a burn-in of one hundred iterations. Fig.~\ref{fig:LC} shows the model fits for nine of the SB2 HBs and Table~\ref{tab:table} reports the median estimates for the argument of periastron, inclination, eccentricity, and period.

\subsection{SB1 Modeling}
\label{sec:SB1S}

\begin{figure*}
    \centering
    \begin{tabular}{ccc}
        \includegraphics[width=0.33\textwidth]{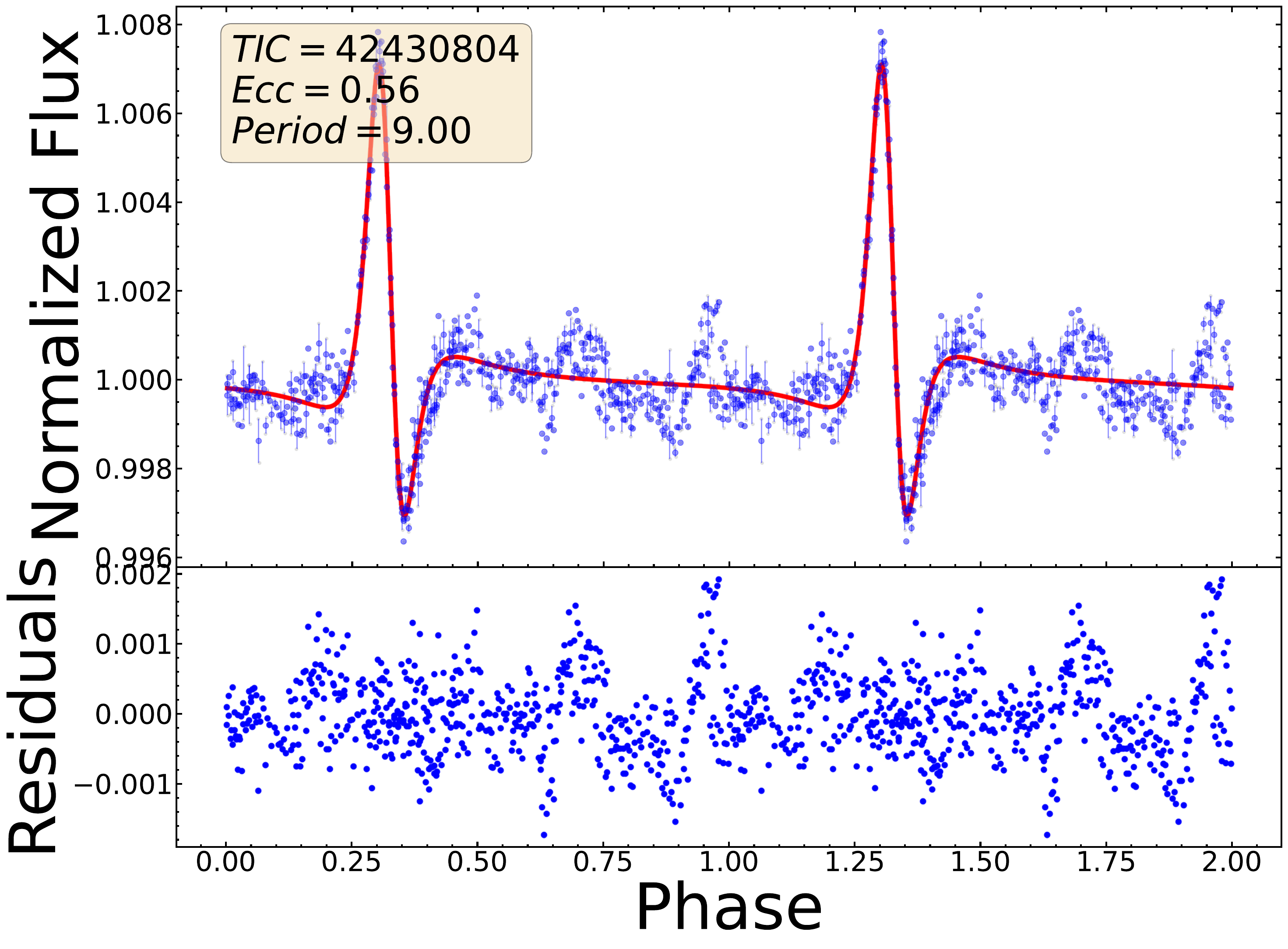} &       
       \includegraphics[width=0.33\textwidth]{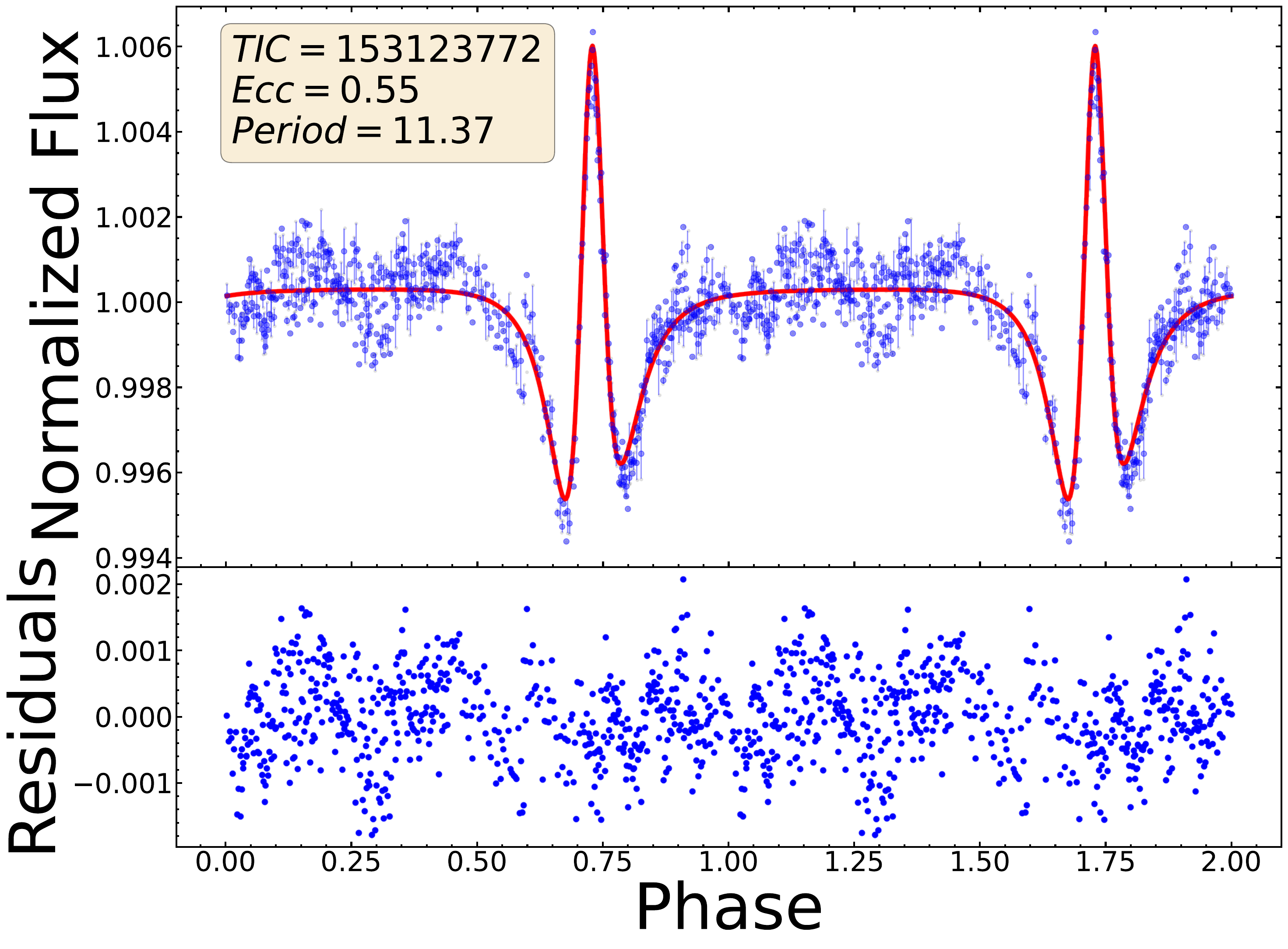} &
        \includegraphics[width=0.33\textwidth]{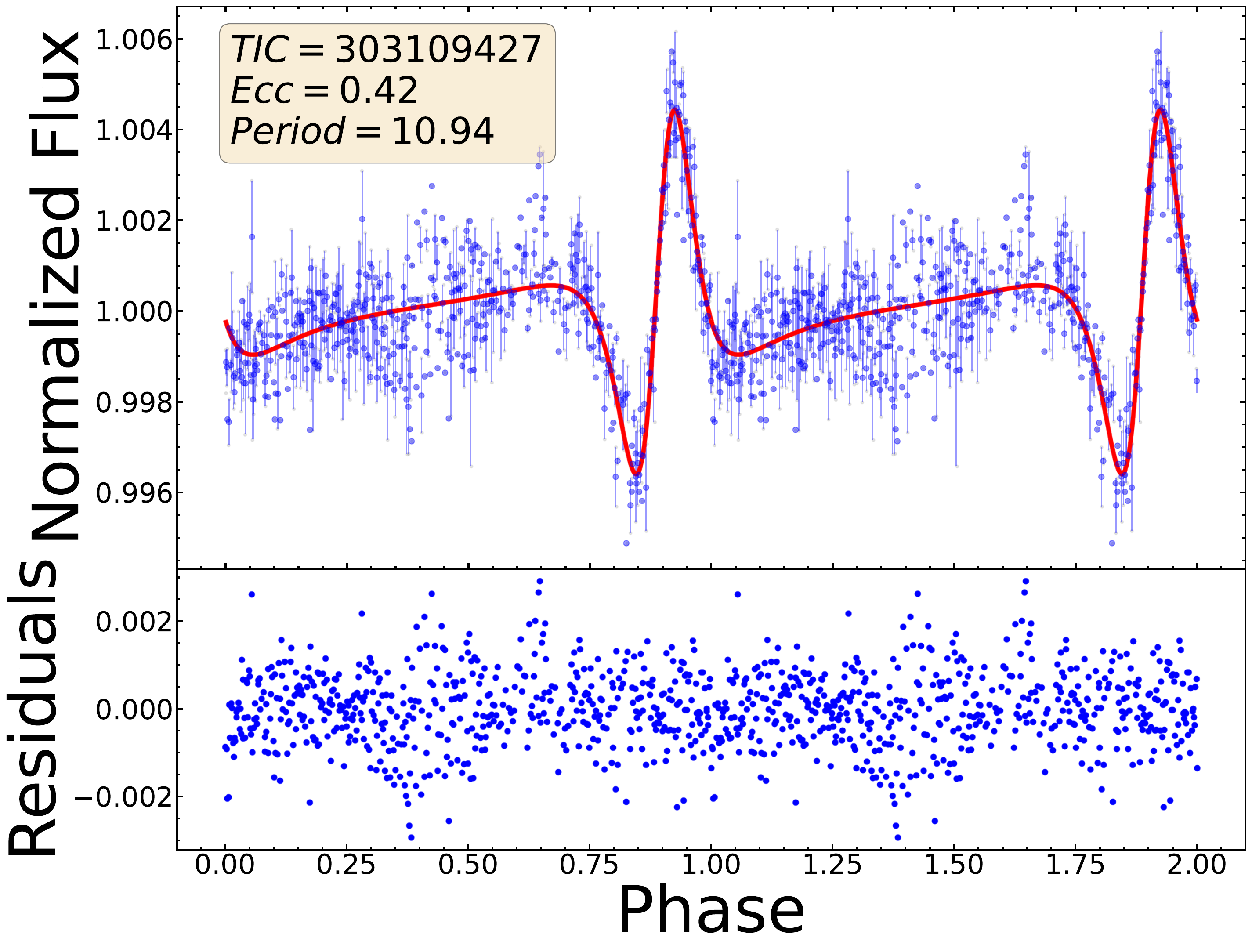} \\
       
       \includegraphics[width=0.33\textwidth]{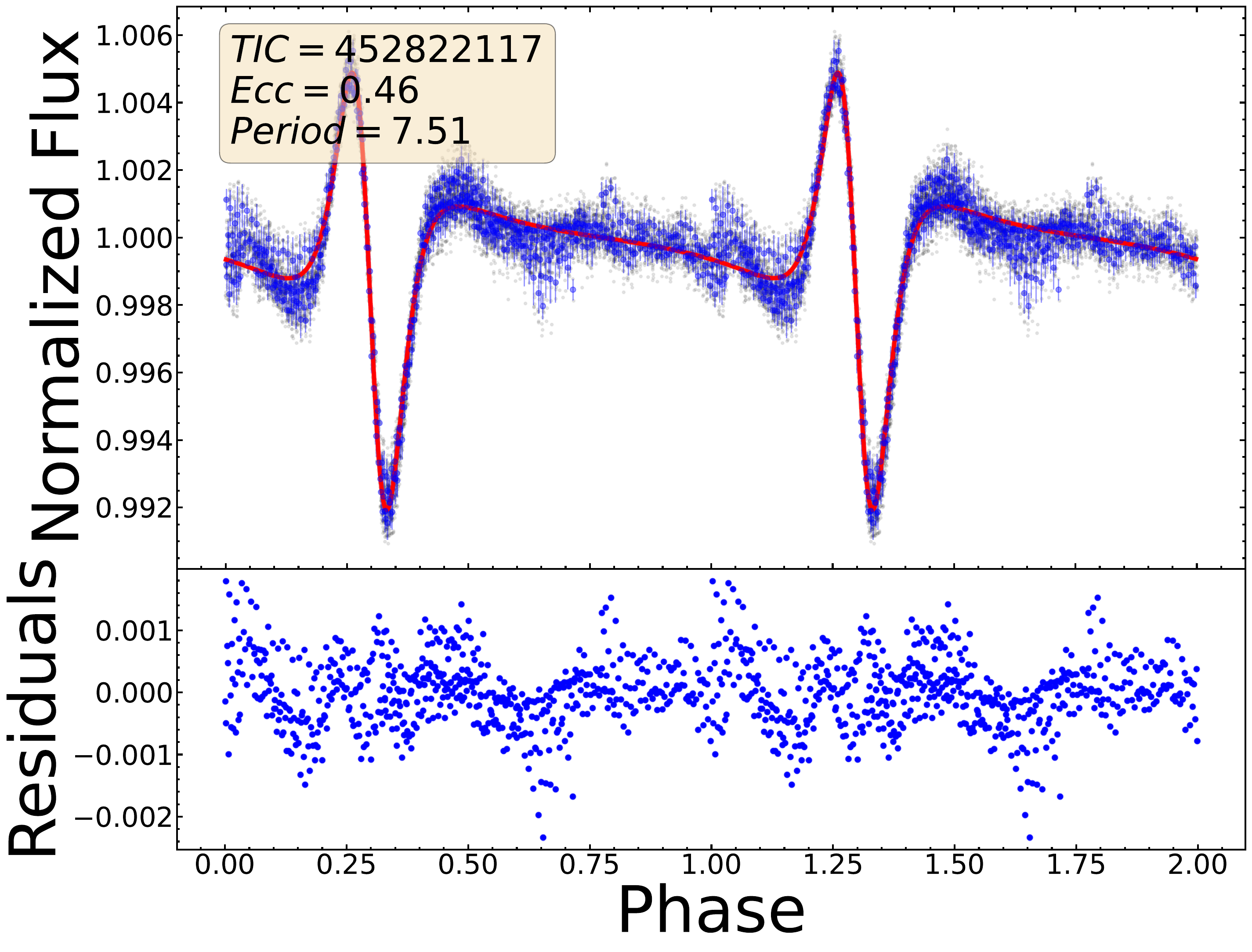} &
	\includegraphics[width=0.33\textwidth]{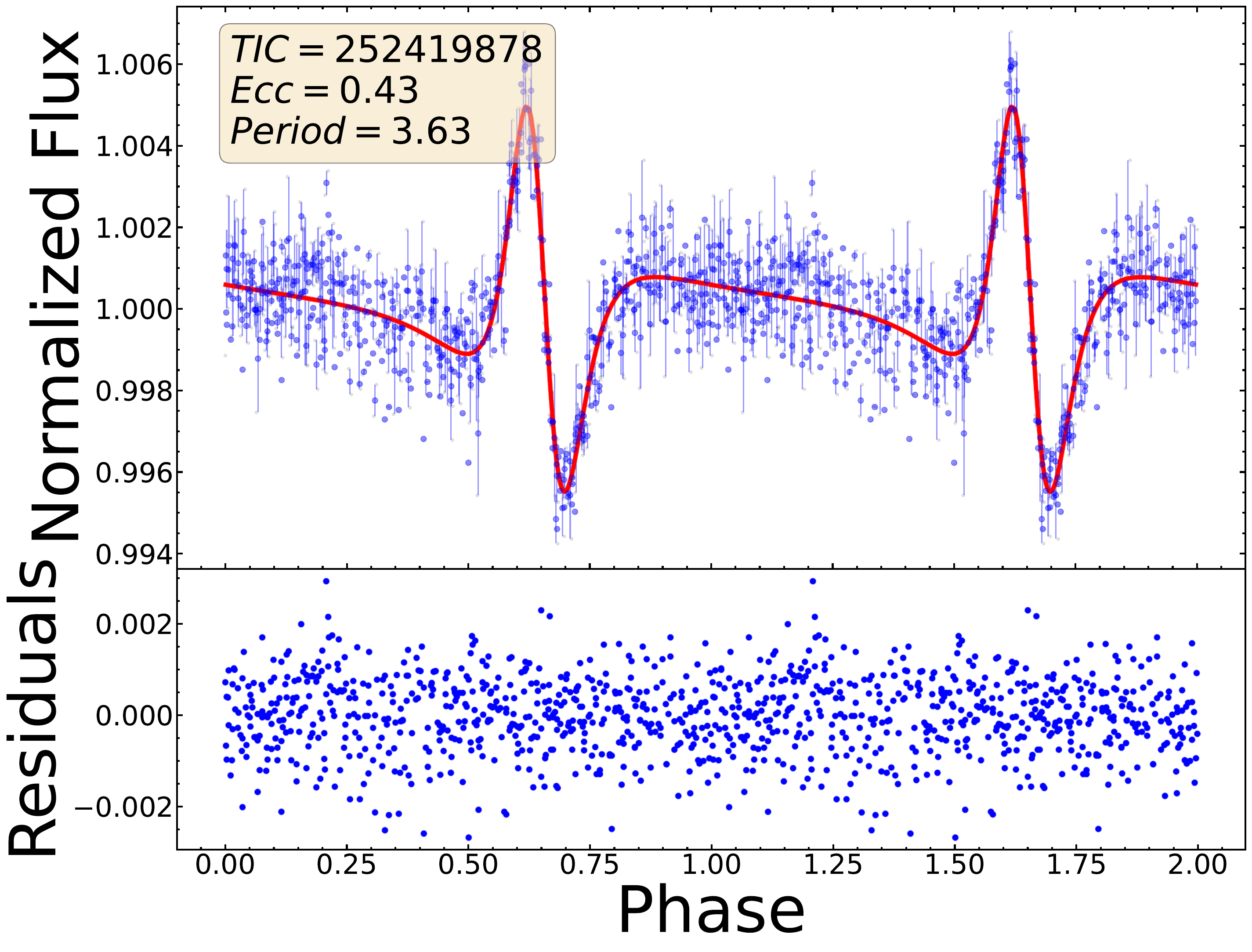} &       
       \includegraphics[width=0.33\textwidth]{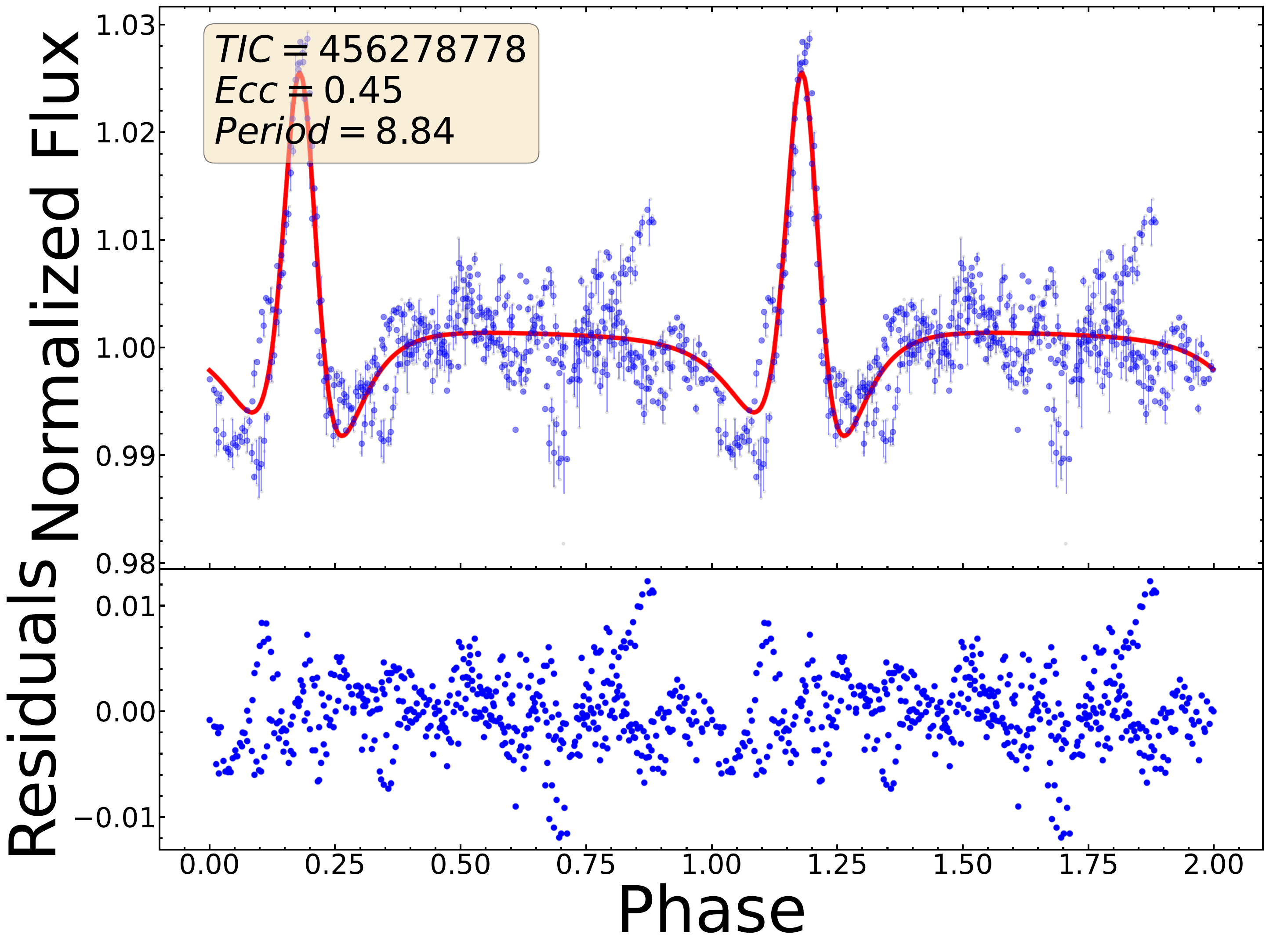} \\
      
        \includegraphics[width=0.33\textwidth]{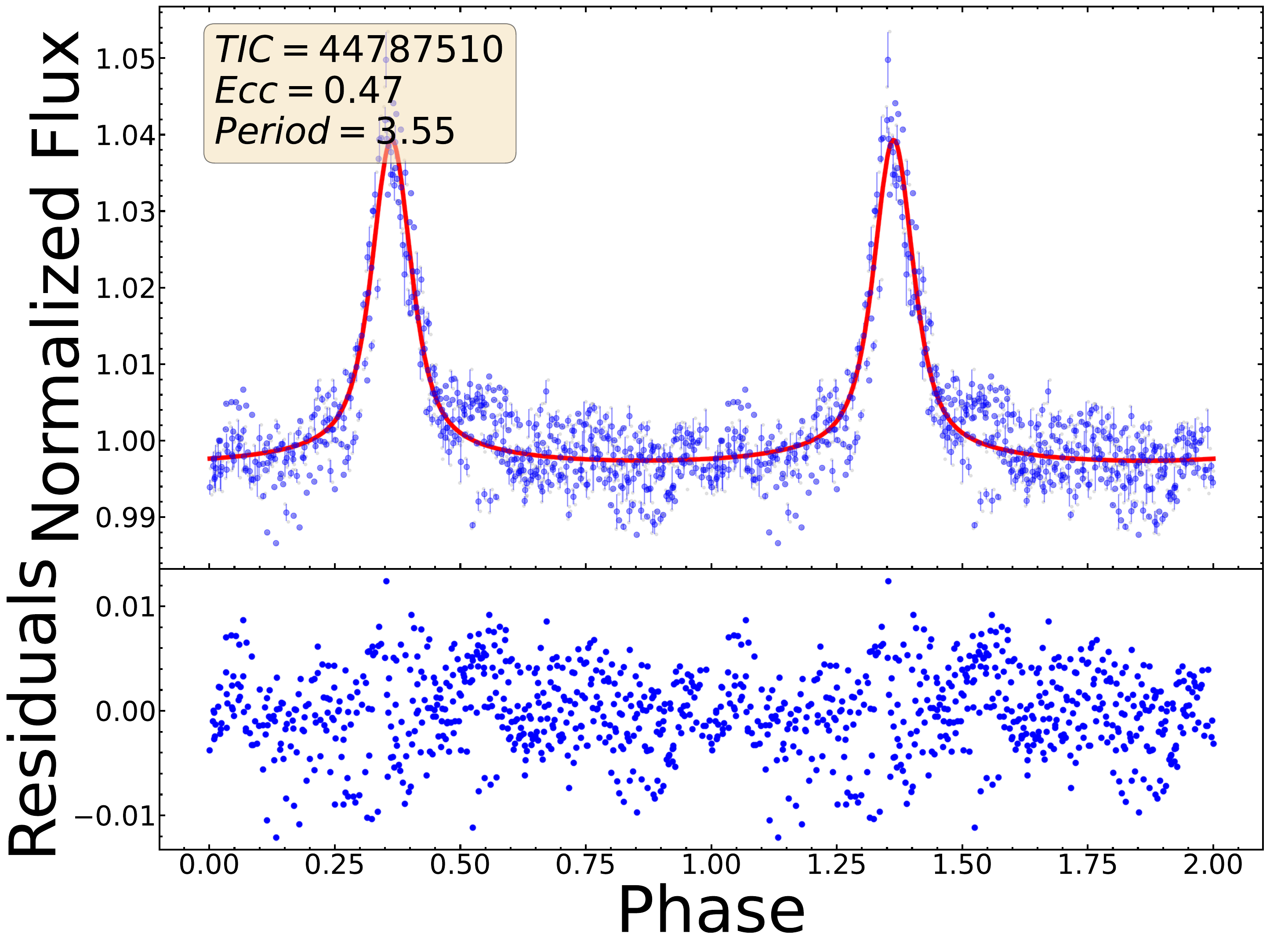} &

       	\includegraphics[width=0.33\textwidth]{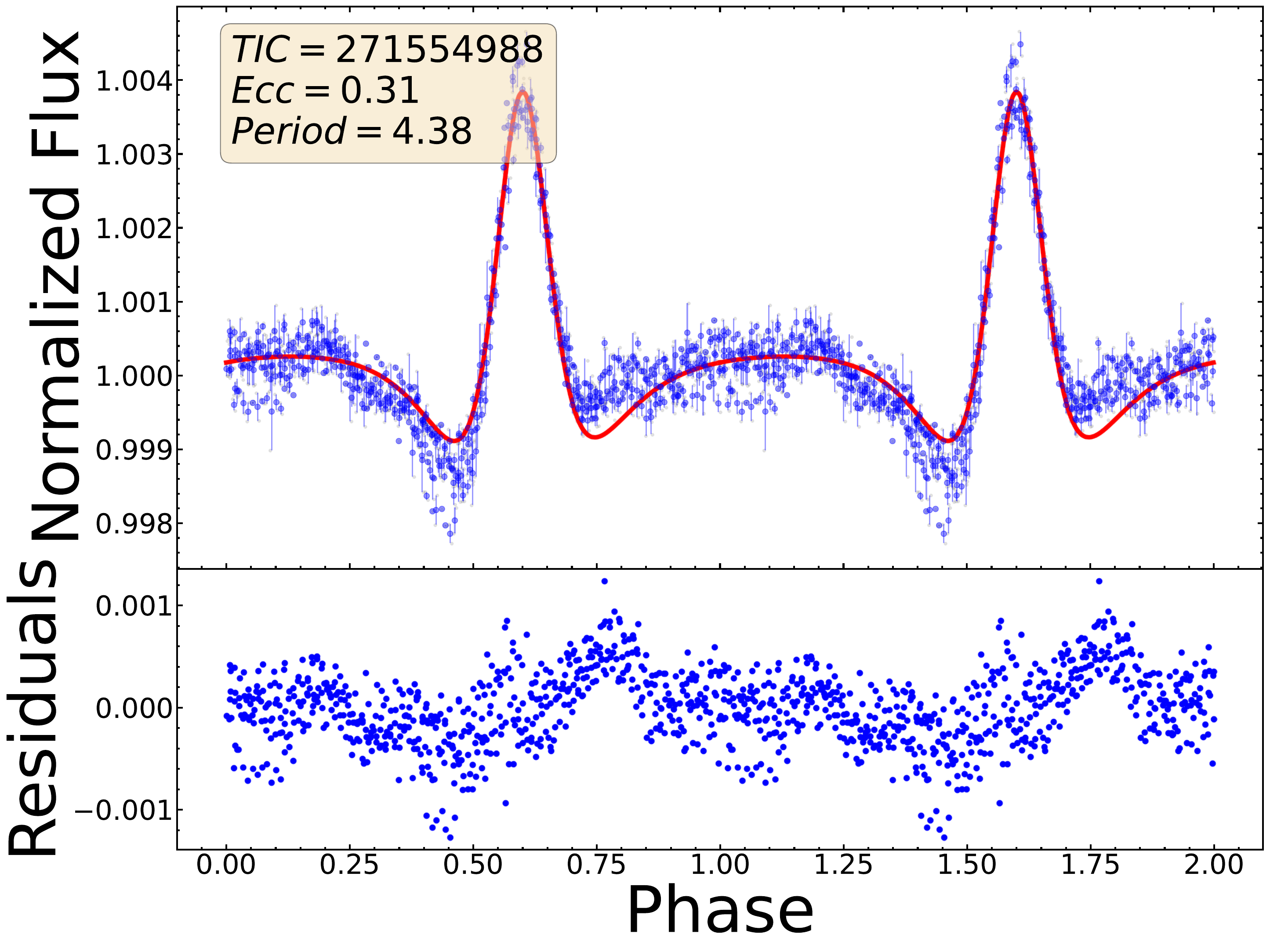} &       
      
       \includegraphics[width=0.33\textwidth]{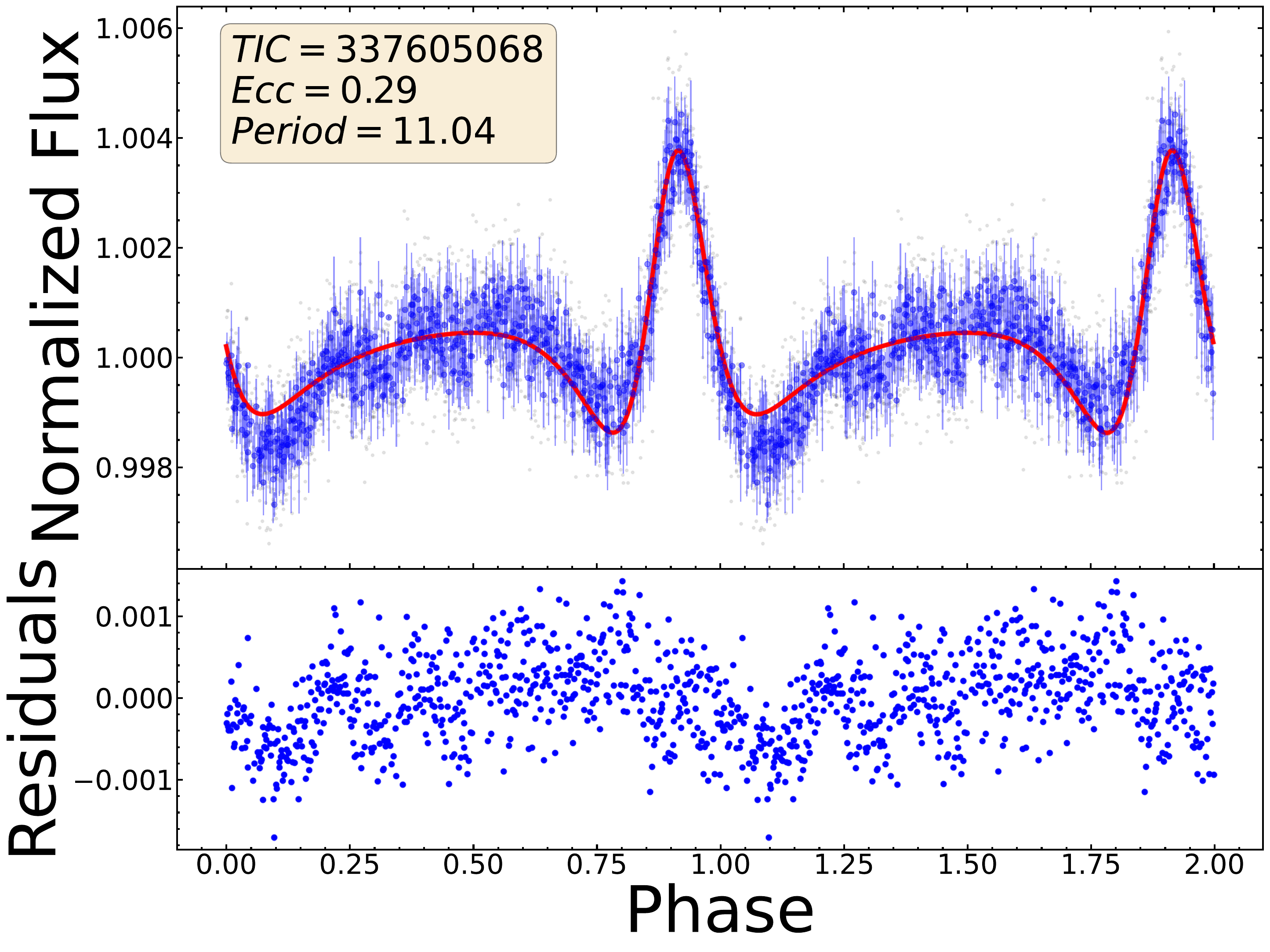} \\
        \includegraphics[width=0.33\textwidth]{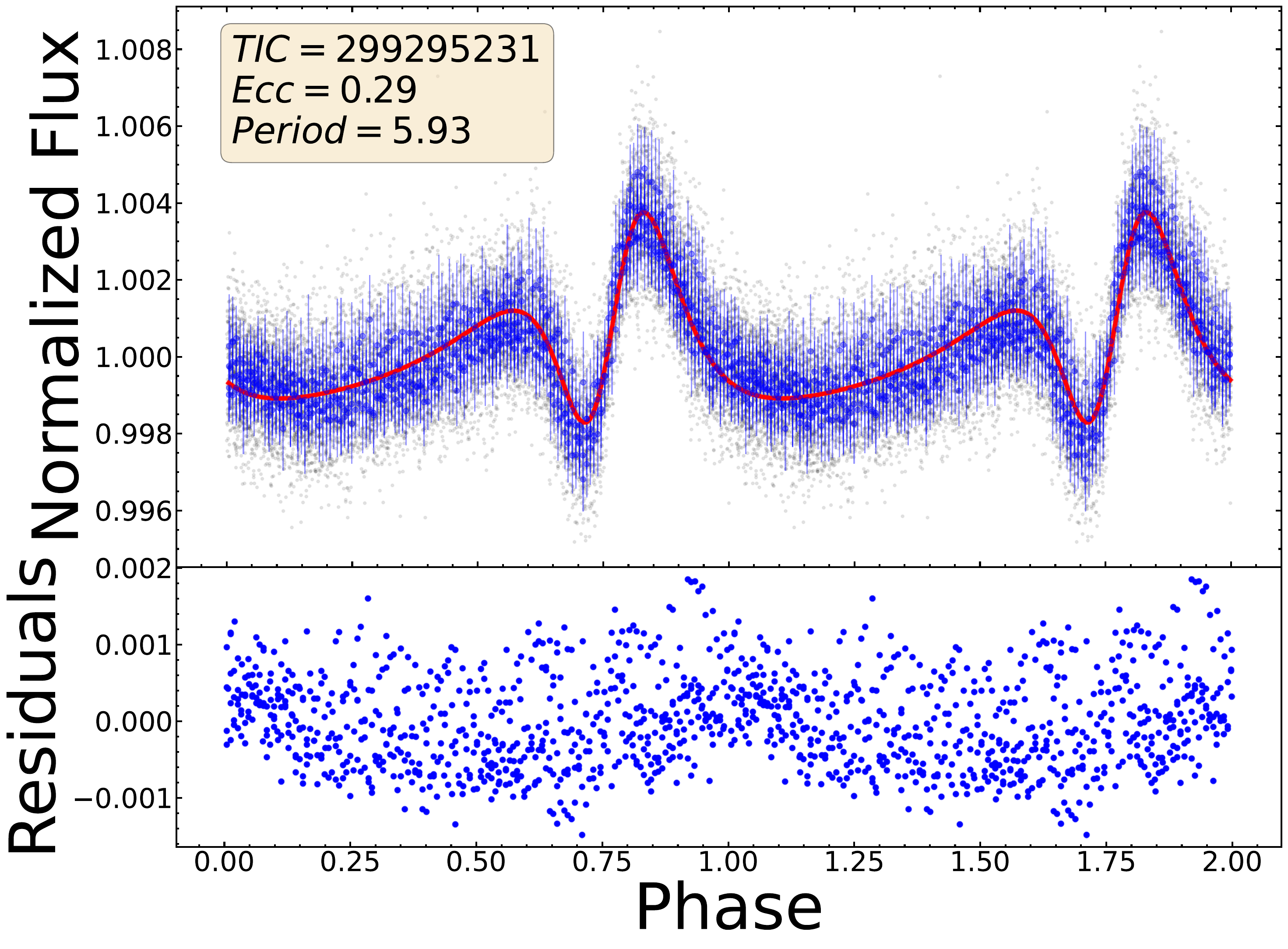} &

       	\includegraphics[width=0.33\textwidth]{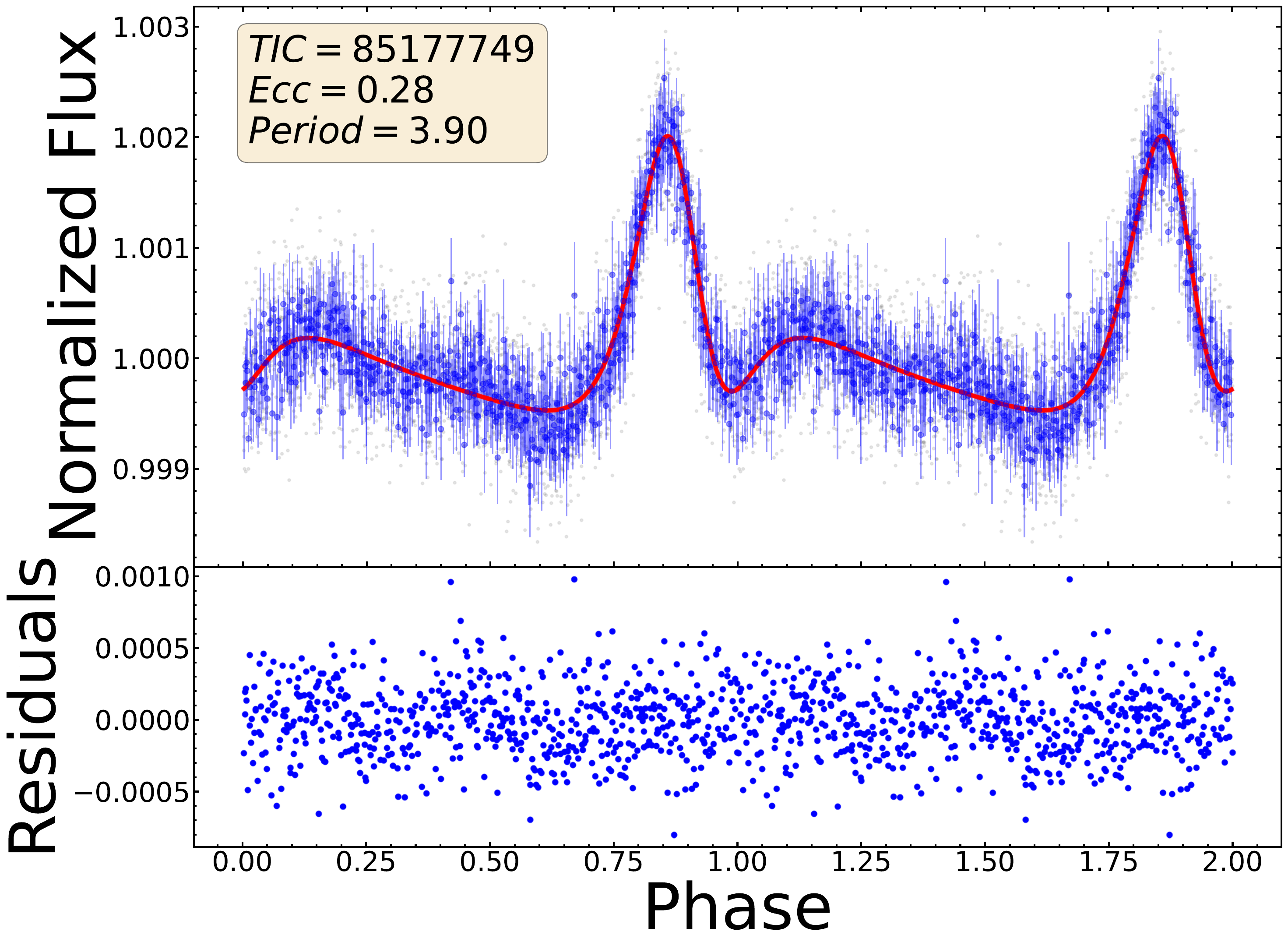} &      
       \includegraphics[width=0.33\textwidth]{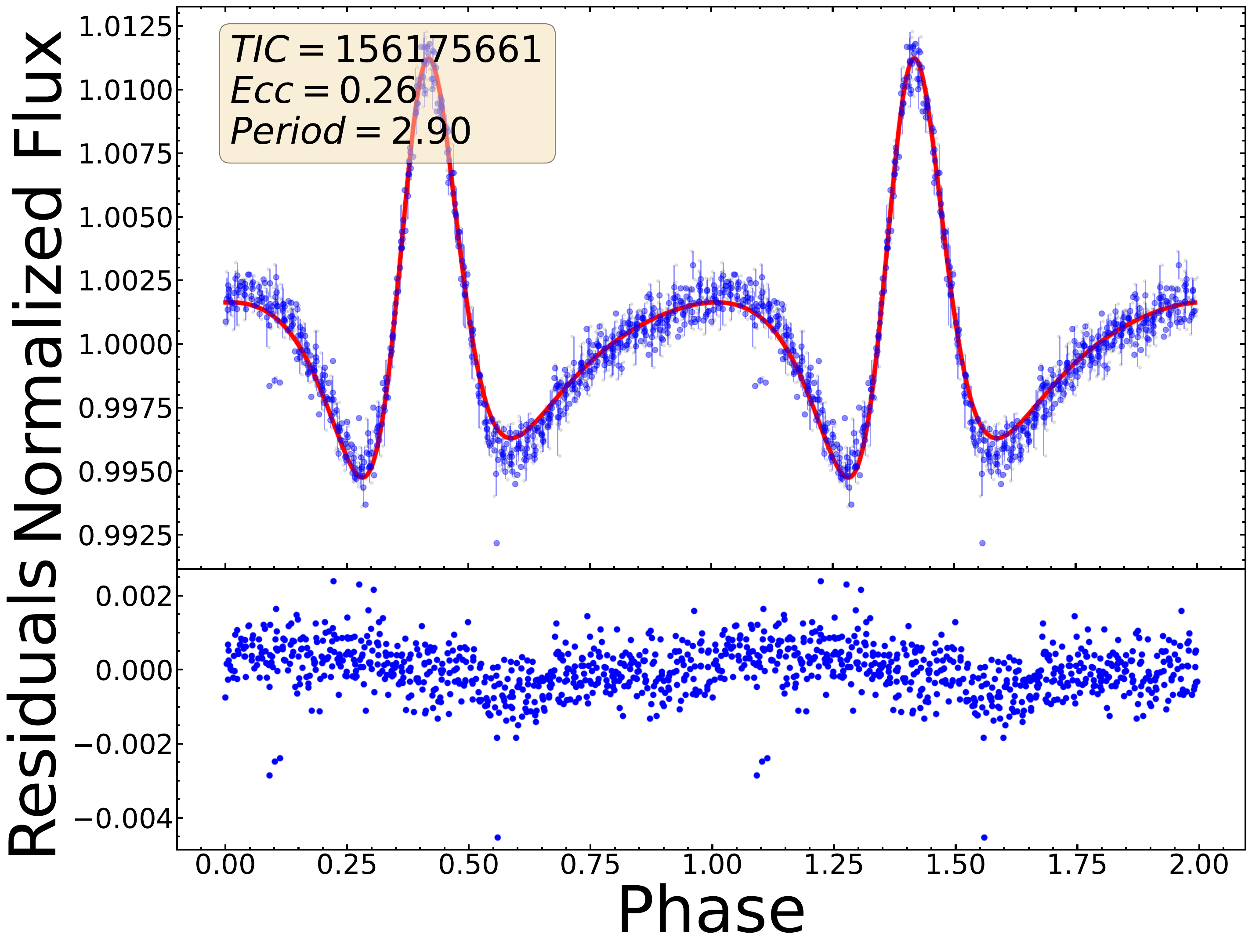} \\

        \includegraphics[width=0.33\textwidth]{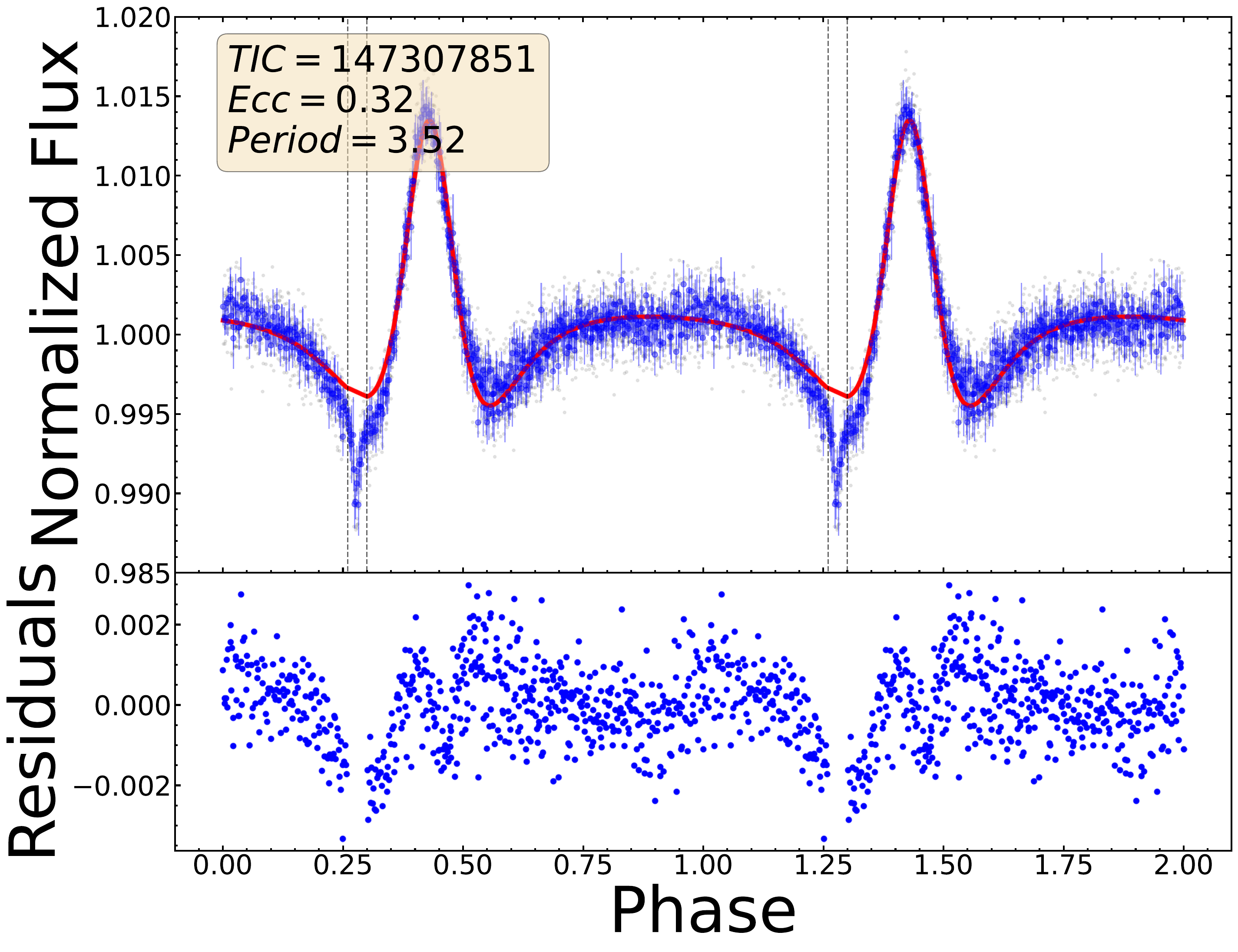} &
	\includegraphics[width=0.33\textwidth]{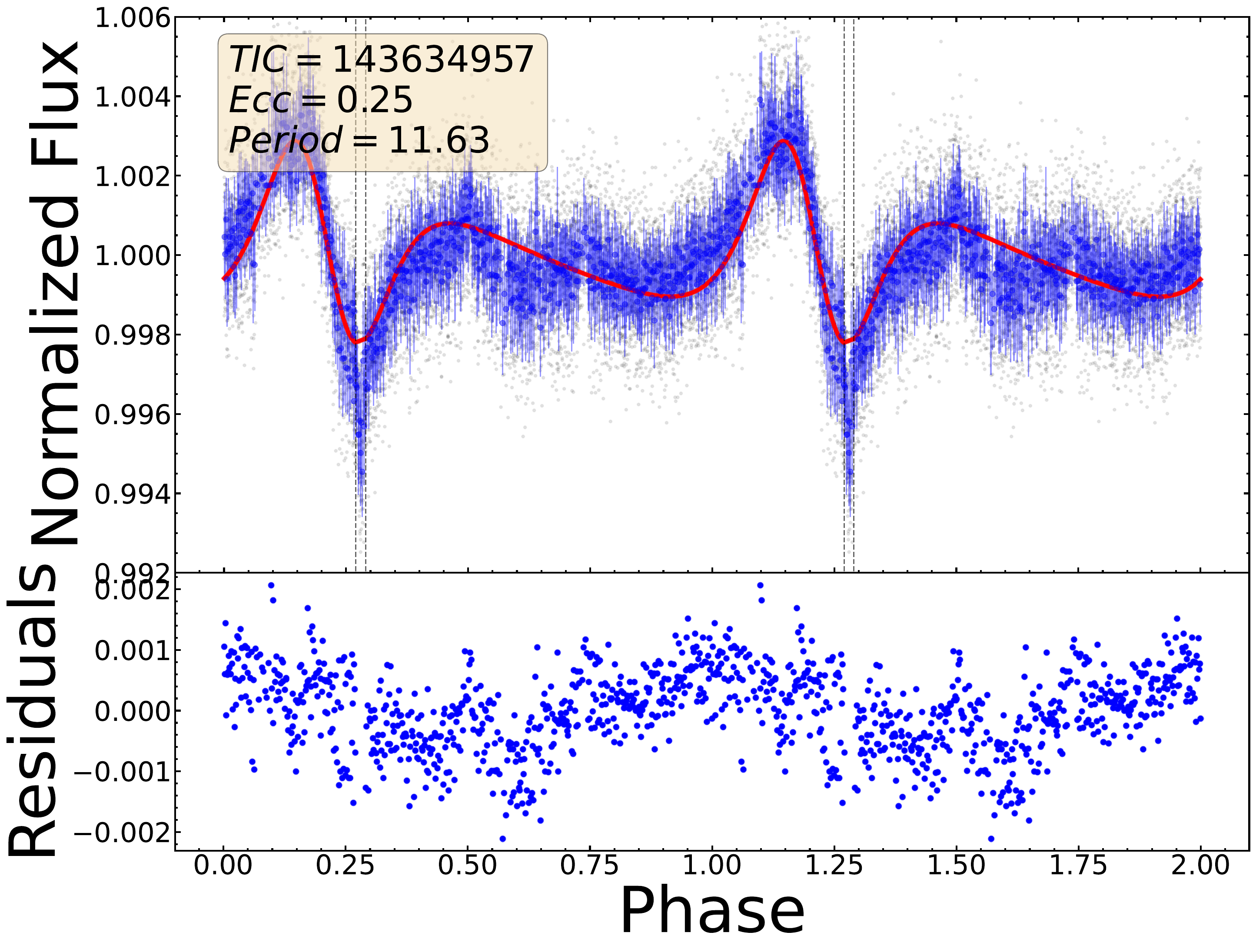} &       
       \includegraphics[width=0.33\textwidth]{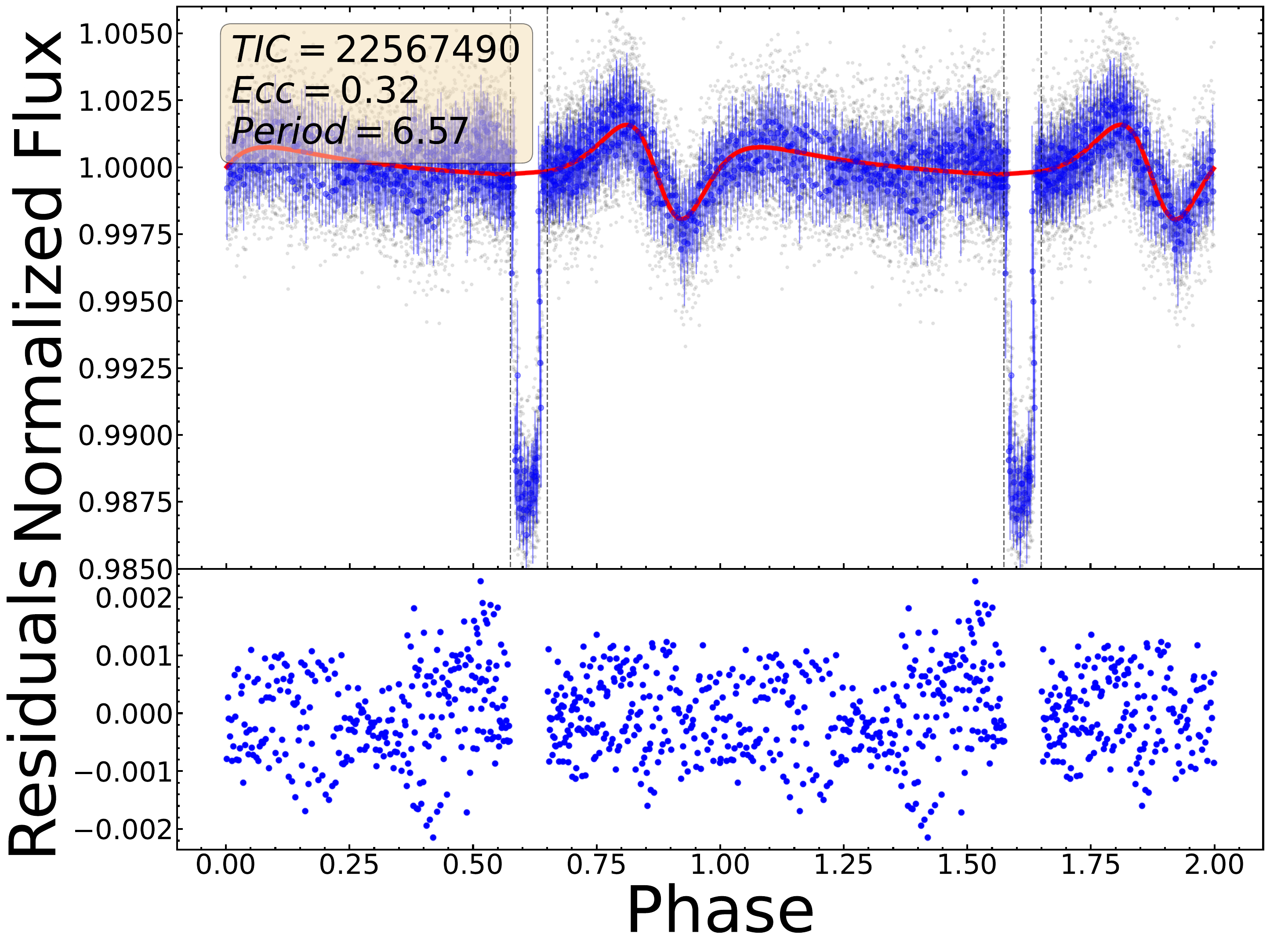} \\
    \end{tabular}
    \caption{Examples of the phase folded light curves of single-line spectroscopic heartbeat binaries with the light curves binned by one hour. The model residuals are shown below each light curve. The text boxes show the \textit{TESS} input catalog (TIC), eccentricity, and period (days) of each target. The bottom row shows three SB1s that also have eclipses between the dashed lines. For the final models, we masked the data in these regions.}
    \label{fig:lcsb2}
\end{figure*}

Since there are 181,529 \textit{Gaia} SB1 orbital solutions, we performed a semi-automated search rather than relying only on visual inspection. We only analyzed the 43,599 systems with \textit{Gaia} orbital periods less than 13 days so that there would be at least two binary orbits in a \textit{TESS} sector. The \textit{TESS} light curves often have features at the start and end of each sector leading to significant numbers of false positives for longer periods. For each light curve sector, we used a LS periodogram to identify periodic variables. We then bin the phase-folded light curves into one hour time-bins, and fit the \citet{Kumar1995} analytic model (Eq.~\ref{eq:Flux}) to the phase-folded light curve using the Trust Region Reflective algorithm. Since the LS periodogram often returns a simple fraction of the orbital period, we do this for $P_{LS}$, $2P_{LS}$, and $3P_{LS,}$ where $P_{LS}$ is the period corresponding to the maximum power in the periodogram.

For each target, we select the sector and period combination that minimizes the reduced $\chi_{HB}^2$ of the Kumar model fit. We then compare this to the $\chi_{line}^2$ for a linear fit to the phase-folded light curves. Fig.~\ref{fig:Final} shows the distribution of $R=\chi_{HB}^2/\chi_{line}^2$ and orbital eccentricities for the 43,599 systems. We use the HBs identified in the SB2 sample to define the selection region ($R<0.5$ and $e>0.15$) shown in Fig.~\ref{fig:Final}. This left 7,314 binaries, which we then visually inspected to select 102 HB candidates. Fig.~\ref{fig:lcsb2} shows examples of the phase-folded light curves with their Kumar model fits, and Table~\ref{tab:table} reports the orbital parameters.

Some of our HB candidates also have primary or secondary eclipses, such as TIC 147307851, TIC 143634957, and TIC 22567490 in Fig.~\ref{fig:lcsb2} (they are flagged in Table~\ref{tab:table}). Since the Kumar model only accounts for tidal distortions, we manually mask the eclipses when necessary and re-fit the phase-folded light curve. In some cases, the revised fit pushes the eccentricity below the 0.15 cutoff used for selection, but we keep these systems in our final sample.

\subsection{The Final Sample}
\label{sec:TFS}

\begin{figure}
	\includegraphics[width=\columnwidth]{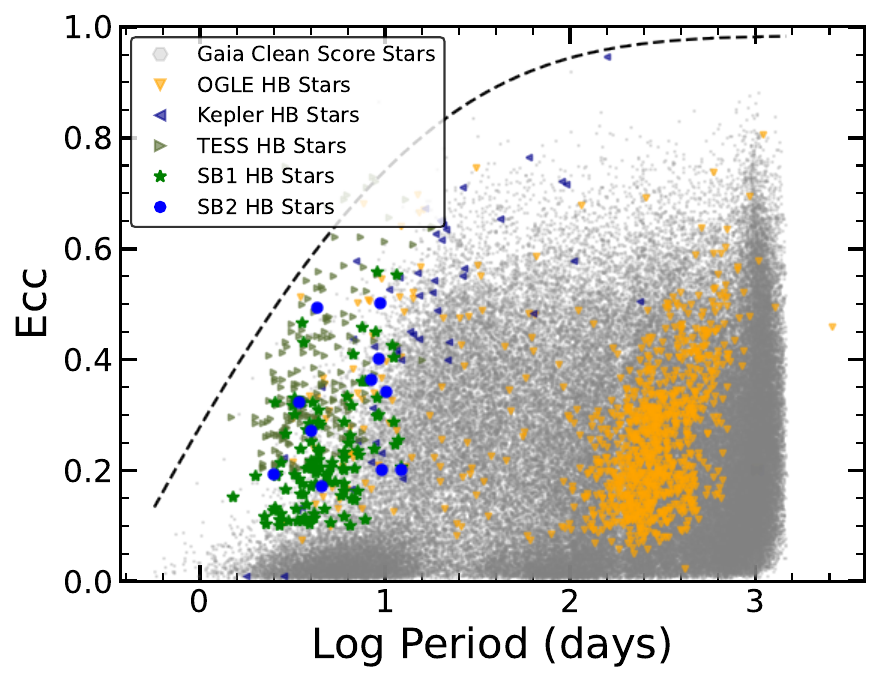}
    \caption{The distribution of heartbeat stars in period and eccentricity, where the ones discussed here are the stars (SB1) and circles (SB2). The grey background is the distribution of all the "clean" \textit{Gaia} SB1s \citep{Bashi2022}. The triangles are the \citet{Wrona2022a, Wrona2022b}, \citet{2023Min-Yu}, and \citet{Sol2025} catalogs of OGLE, \textit{Kepler}, and \textit{TESS} HBs. The curve is the envelope outside of which orbits will rapidly circularize \citep{Mazeh2008}.}
    \label{fig:Mozeh}
\end{figure}

\begin{figure}
    \centering
    \begin{tabular}{c}
	\includegraphics[width=\columnwidth]{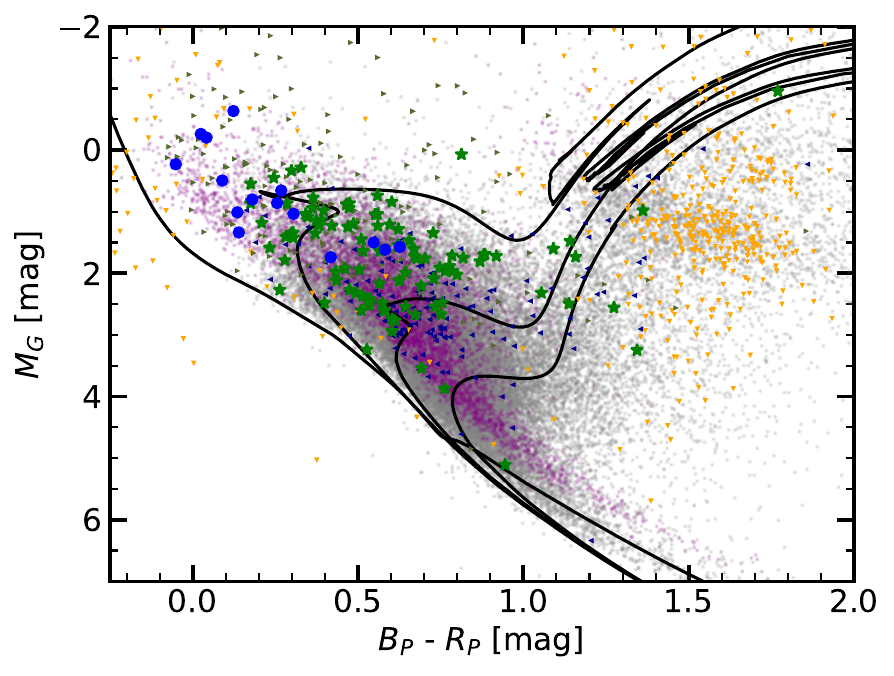} \\
    \end{tabular}
    \caption{Colour Magnitude Diagram (CMD) of the \textit{Gaia} SB1 and SB2 systems (grey and purple background), the SB1 (stars) and SB2 (dots) HBs detected here, and the HB systems (triangles) from \citet{Wrona2022a, Wrona2022b}, \citet{Shporer2016}, \citet{2023Min-Yu}, \citet{2017Dimitrov}, \citet{Kirk2016}, \citet{SK2021}, and \citet{Sol2025}. For the OGLE Magellanic Cloud HBs we used the \citet{2021Skowron} reddening map and the distance and extinction assumptions from \citet{2025MacLeod}. The 0.5, 1, 5, 10 Gyr isochrones are solar metallically MIST isochrones \citep{2016Dotter, 2016Choi, 2011Paxton, 2013Paxton, 2015Paxton, 2018Paxton}.}
    \label{fig:CMD}
\end{figure}

\begin{figure}
    \centering
    \begin{tabular}{cc}
	\includegraphics[width=0.5\columnwidth]{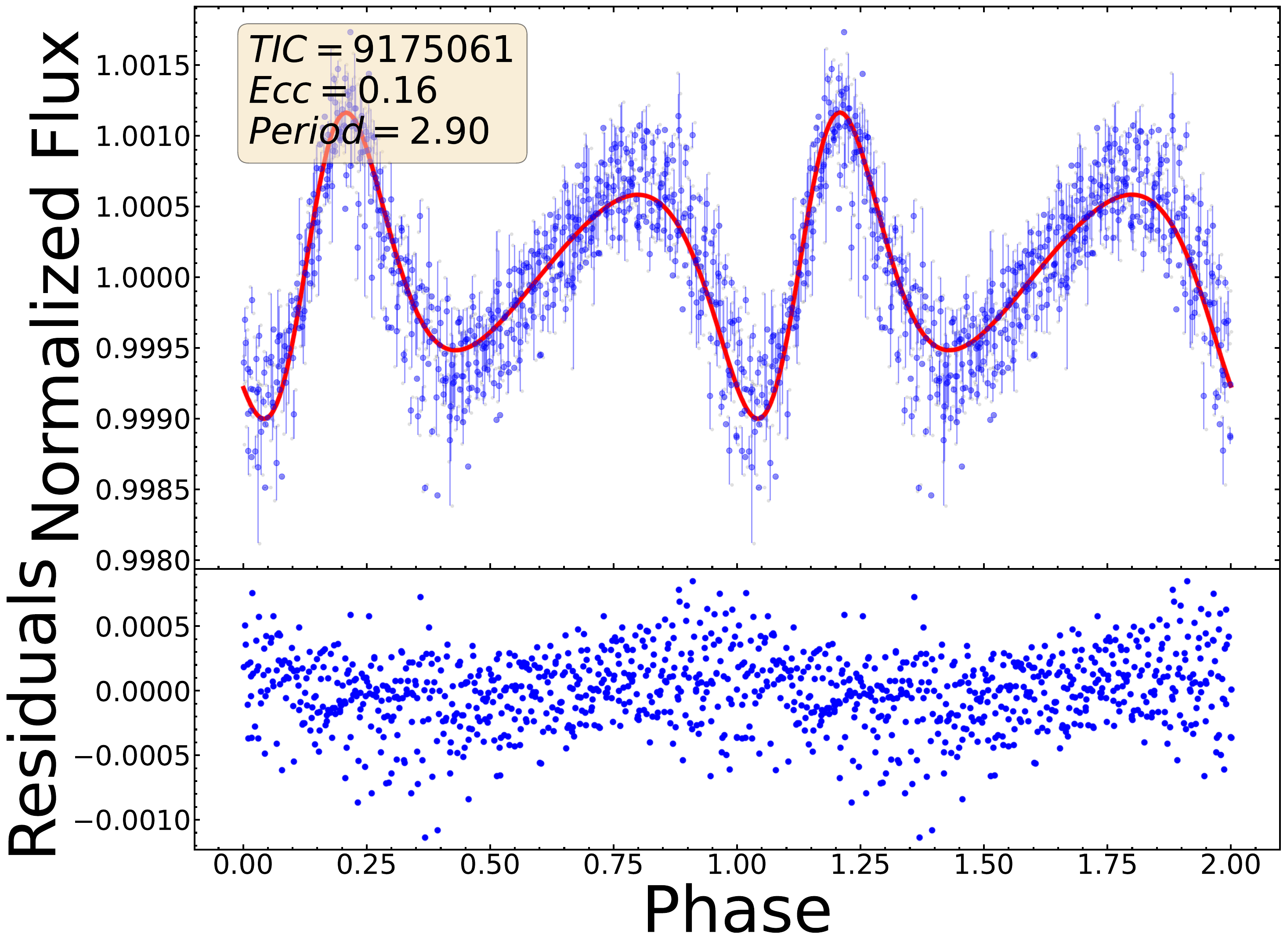} &
    \includegraphics[width=0.5\columnwidth]{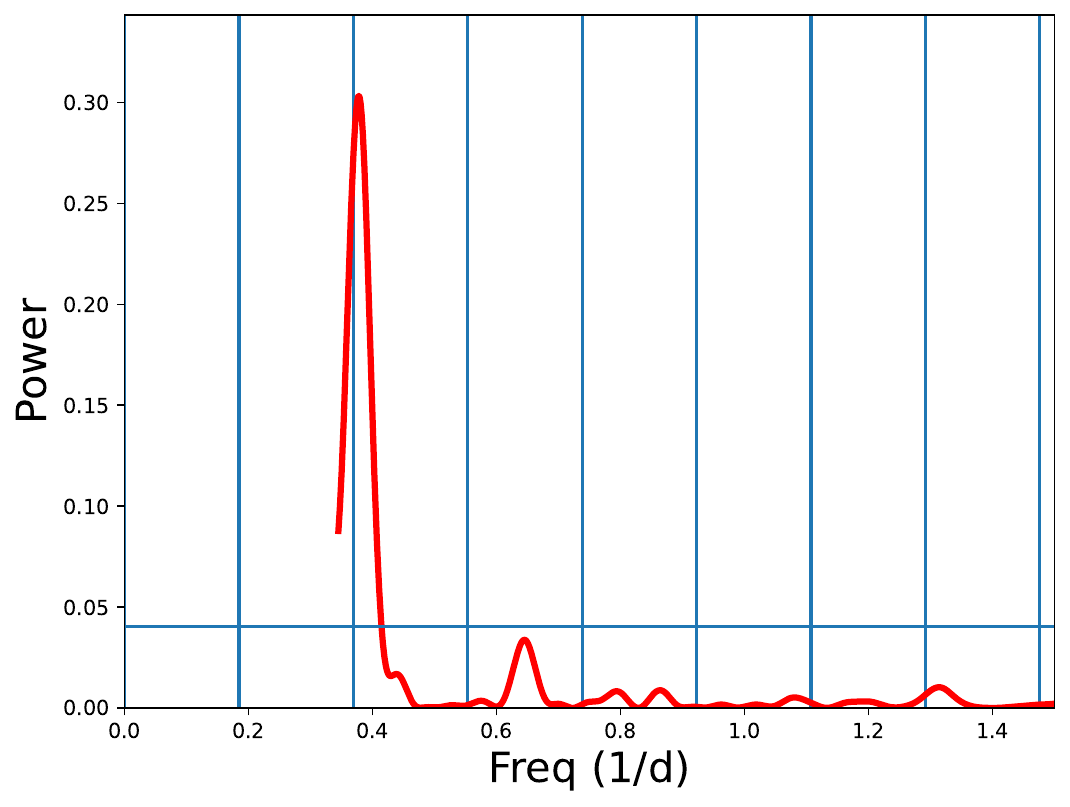} \\
    \includegraphics[width=0.5\columnwidth]{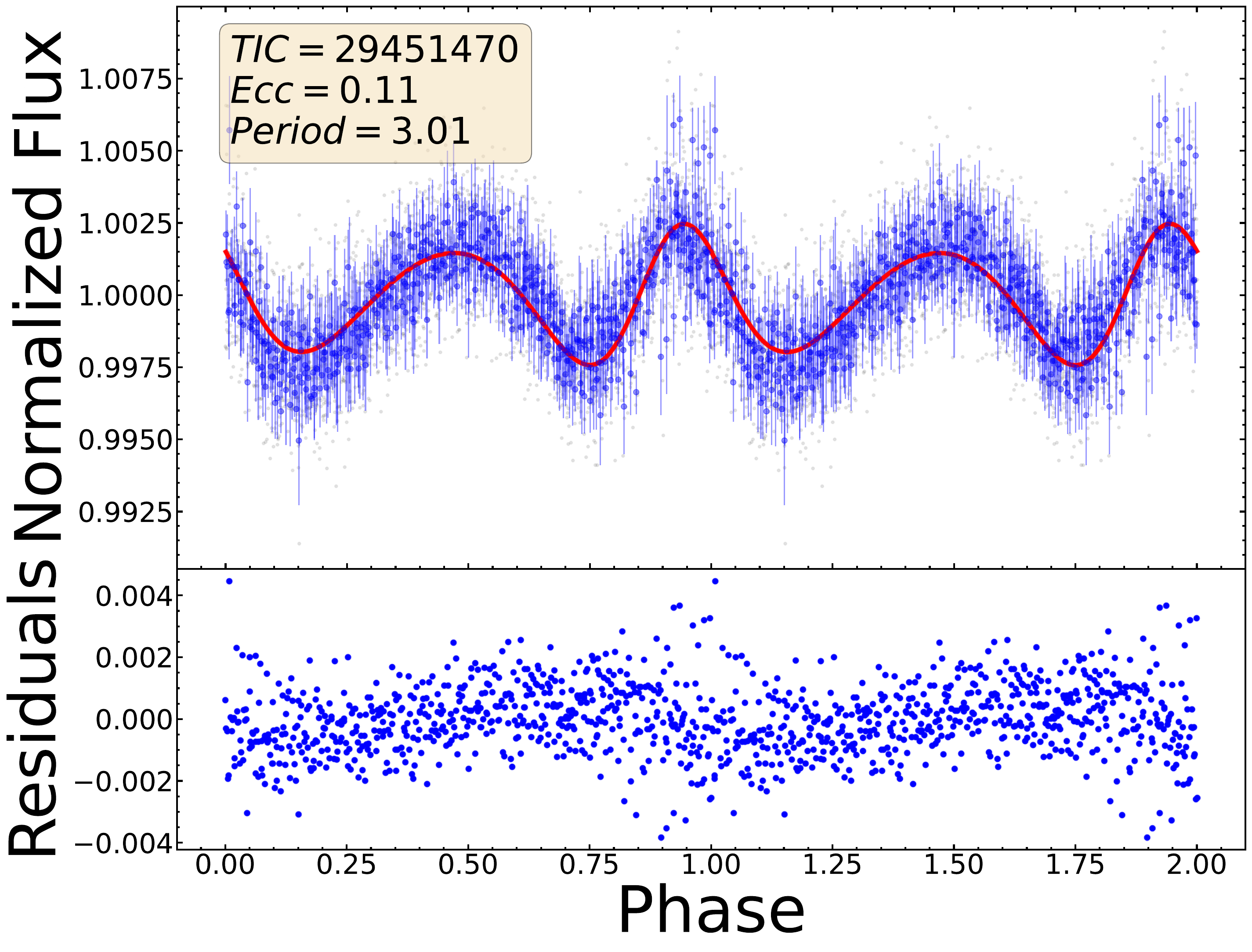} &
    \includegraphics[width=0.5\columnwidth]{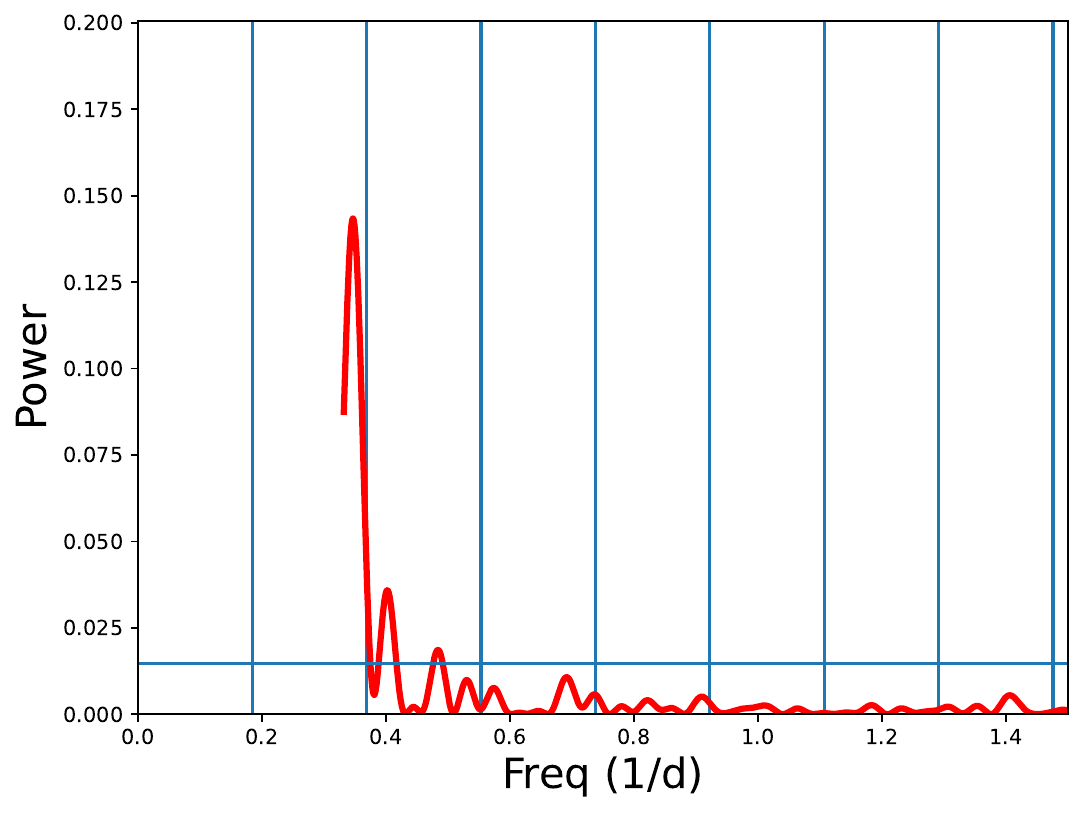} \\
    \includegraphics[width=0.5\columnwidth]{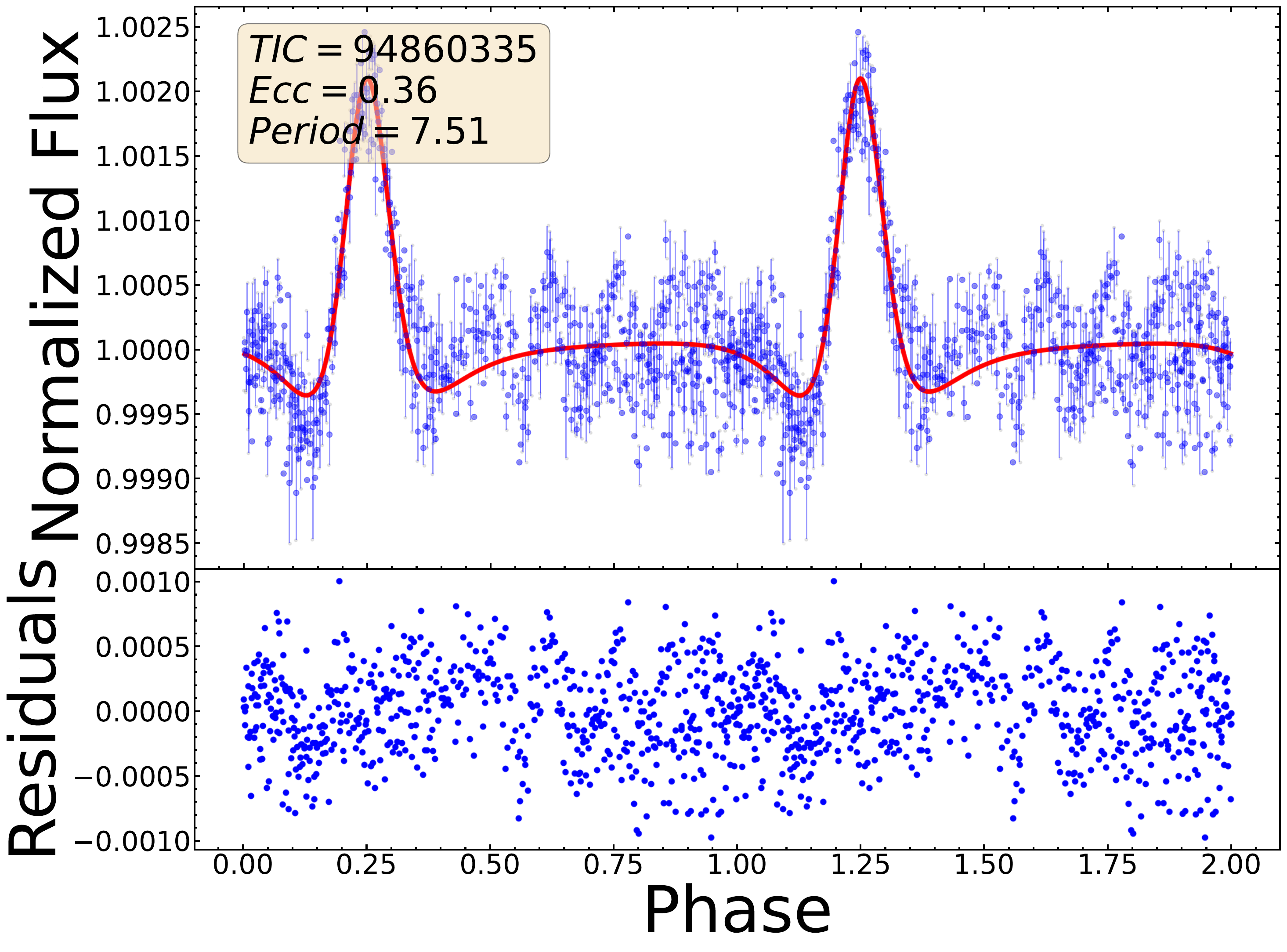} &
    \includegraphics[width=0.5\columnwidth]{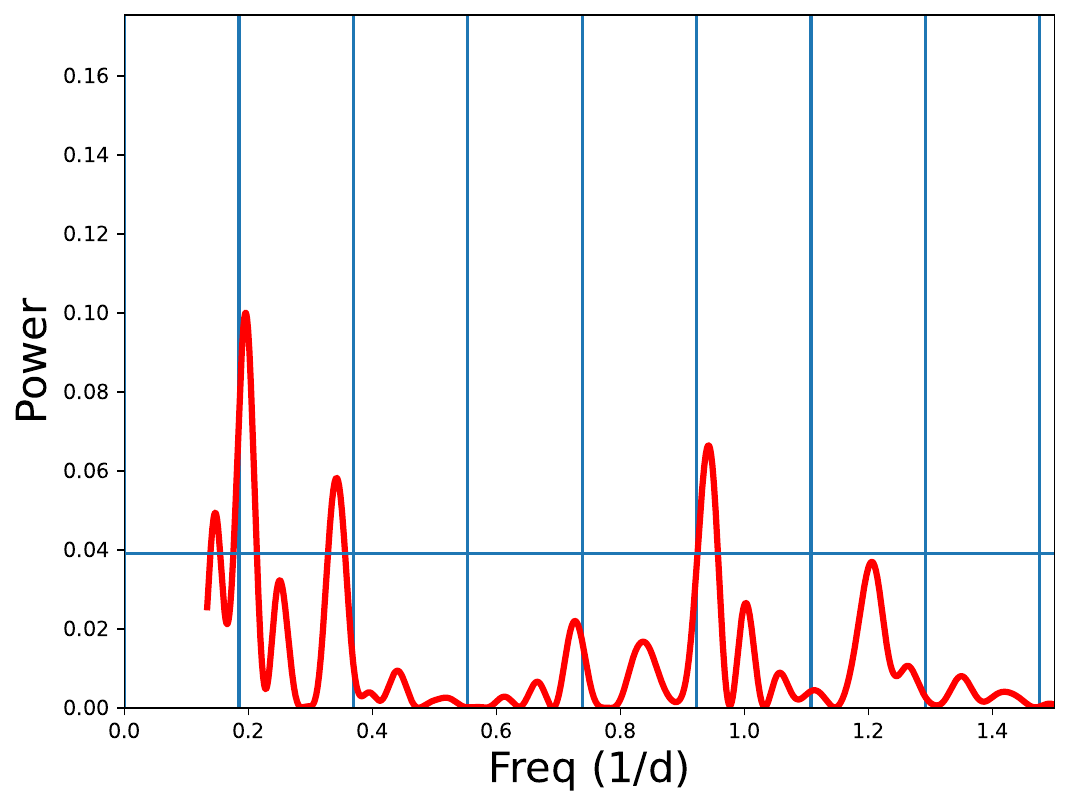} \\
    \end{tabular}
    \caption{Examples of targets with tidally excited oscillations. The left panel shows the phase folded binned light curves, the model and the model residual. The text boxes show the \textit{TESS} input catalog (TIC), eccentricity, and period (days) of each target. The right panel shows the periodogram of the residuals with vertical lines at orbital harmonic frequencies ($n/P$) and a horizontal line at a false alarm probability of $10^{-5}$.}
    \label{fig:TEO}
\end{figure}

Fig.~\ref{fig:Mozeh} shows the orbital periods and eccentricities of the 112 HBs detected in the \textit{Gaia} SB1 and SB2 samples. Of these 112 HBs, four were found by \citet{Sol2025} (TIC 252588526, TIC 271554988, TIC 344586348, and TIC 370209445). We found no matches to the other comparison samples used below. The orbital periods range from 1.51 to 12.17 days, with a median of 4.36 days. The eccentricities range from 0.10 to 0.57 (TIC 42430804), with a median of 0.20. All of our targets fall below the period-eccentricity envelope expected from tidal circularization \citep{Mazeh2008}. Fig.~\ref{fig:Mozeh} also shows the HBs from \citet{Wrona2022a, Wrona2022b}, \citet{2023Min-Yu}, and  \citet{Sol2025}. As compared to the HBs identified in \citet{Wrona2022a, Wrona2022b}, our systems have much shorter periods than their median value of 265.3 days. The median eccentricities are similar (0.24 for OGLE). Our period cut will exclude giants while the OGLE magnitude limits favour giants given the distance to the Magellanic Clouds.

Fig.~\ref{fig:CMD} shows the stars on an extinction-corrected \textit{Gaia} CMD. We use distances from \citet{Bailer-Jones2021} and extinctions from the \texttt{mwdust} three-dimensional dust map \citep{Bovy2016, Drimmel2003, 2006Marshall, 2019Green}. The candidates have good parallaxes ($\varpi/\sigma_{\varpi}>10$) and none are highly extincted ($A_V<2.0$ mag), so their CMD locations should be reliable. As compared to the full sample of \textit{Gaia} spectroscopic binaries, the HBs tend to be higher on the MS (median $M_G=1.64$). Fig.~\ref{fig:CMD} also includes 478 HBs from OGLE \citep{Wrona2022a, Wrona2022b}, 160 from \textit{Kepler} \citep{Shporer2016, 2023Min-Yu, 2017Dimitrov, Kirk2016}, and 179 from \textit{TESS} \citep{SK2021, Sol2025}.

We used periodograms of the residuals of the \citet{Kumar1995} models, to search for TEOs. We list the systems with periodogram powers larger than a false alarm level of $10^{-5}$ at a simple harmonic $n/P$ of the orbit in Table~\ref{tab:table}. Ten of the 102 SB1 HB systems show TEOs, Fig.~\ref{fig:TEO} shows three examples. Similar to \citet{Li2024b}, our TEOs appear at low orbital harmonics with a median value of $n=3$ and a maximum at $n=16$ for TIC 173561516. The TEOs can be used to study the internal structure of these stars.

\section{\textit{Gaia} Comparison}
\label{sec:GC}

\begin{figure}
    \centering
    \begin{tabular}{c}
	\includegraphics[width=\columnwidth]{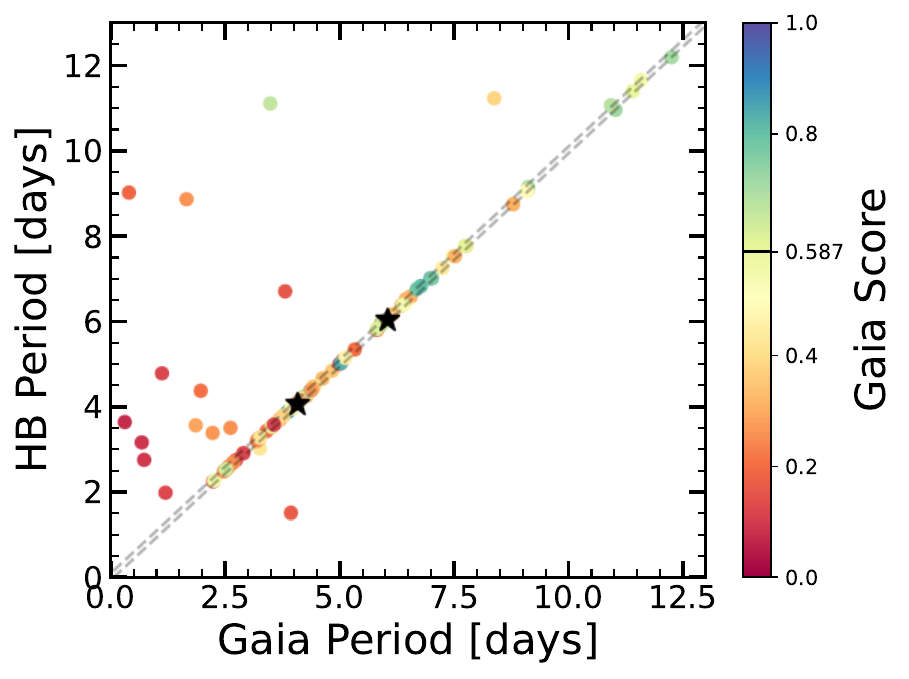} \\
	\includegraphics[width=\columnwidth]{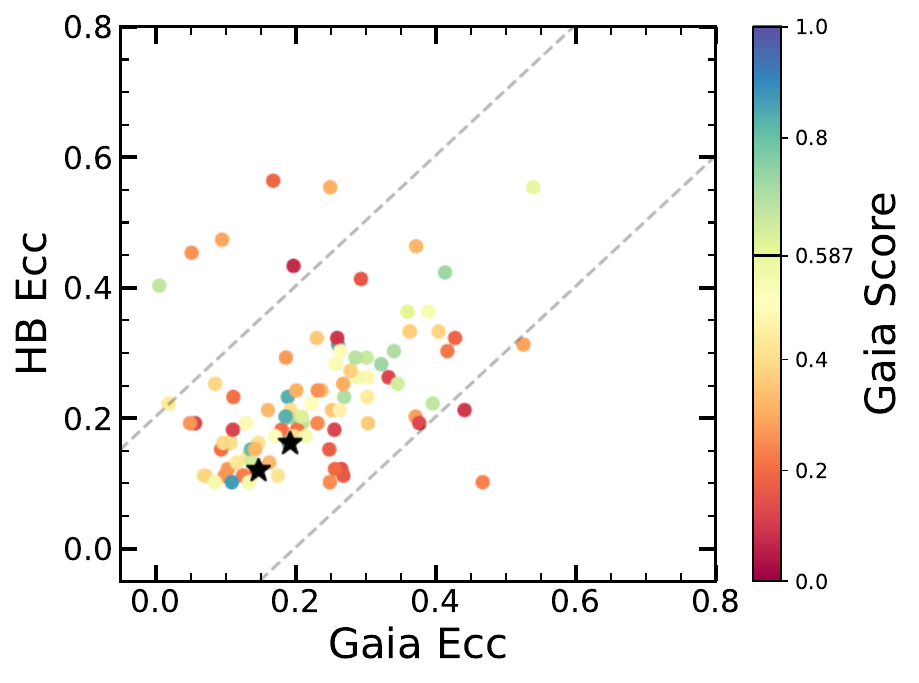} \\
	\includegraphics[width=\columnwidth]{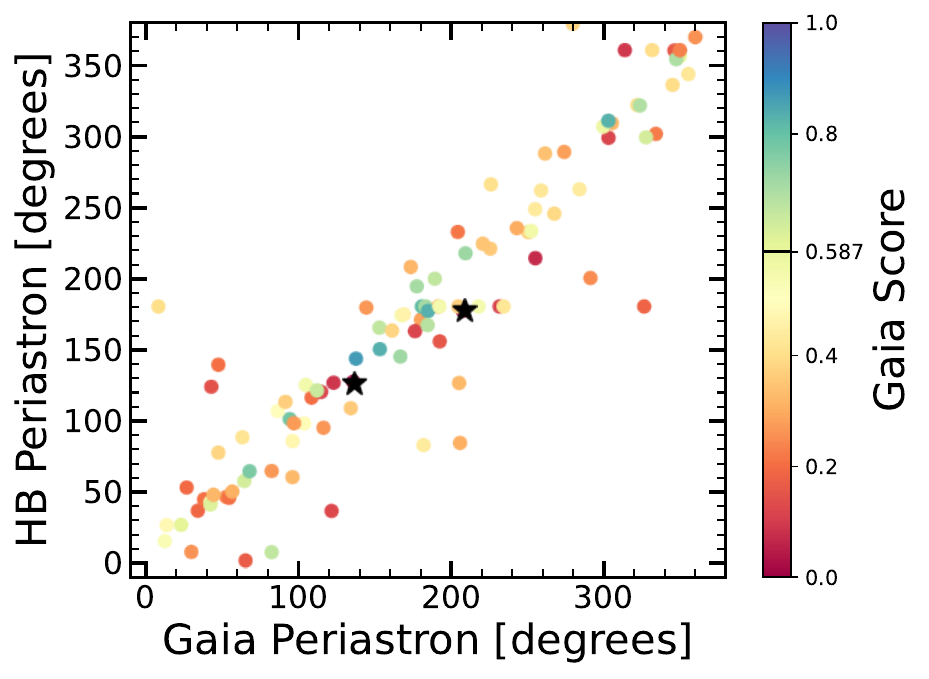} \\
    \end{tabular}
    \caption{Comparisons of the \textit{Gaia} SB1 orbital parameters and those from the HB fits in Table~\ref{tab:table}. The points are coloured by the \textit{Gaia} score from \citet{Bashi2022}, centered on their score criterion of $S<0.587$ for good orbital solutions (colour bar). The dashed lines are the range we consider a reasonable match between the values. Two of our HBs, marked with stars, do not have scores.} 
    \label{fig:COMP}
\end{figure}

Fig.~\ref{fig:COMP} compares the \textit{Gaia} spectroscopic orbit solutions and HB fit orbital periods, eccentricities, and arguments of periastron. We were particularly interested in the SB2s because agreement between the orbital parameters means that the \textit{Gaia} velocity semi-amplitudes can be combined with the photometric variability to measure masses and radii \citep{Rowan2023}. Unfortunately, only 2 of the 10 SB2 HBs have orbital periods and eccentricities that are in good agreement. It is difficult to pinpoint why an individual \textit{Gaia} orbital solution is incorrect, since the epoch RVs are not included in \textit{Gaia} DR3, but it is likely due to the large number of degenerate periods that are possible when fitting sparse RVs measured over a long time baseline to short period orbits. Furthermore, most of the SB2 HBs are B/A stars on the upper MS (Fig.~\ref{fig:CMD}), and the 846--870nm wavelength range of the \textit{Gaia} RVS instrument was primarily designed for cool stars \citep{2018Cropper}. This is one reason that the RVs for hot stars only became available with DR3 \citep{Blomme2023}.

The SB1 HBs have considerably more accurate \textit{Gaia} orbits, with 85$\%$ of the orbital periods and eccentricities agreeing with the light curve solution. \citet{Bashi2022} introduced a “score” statistic ranging from $0\leq S\leq 1$ to quantify the accuracy of \textit{Gaia} SB1 orbital solutions. Of the 87 matching SB1 HB systems, 73 have scores below the $S=0.587$ cutoff recommended by \citet{Bashi2022} for the selection of a “clean” SB1 sample. Despite this, the score statistic seems to be a limited indicator of the \textit{Gaia} orbit quality, as almost all of the targets with incorrect \textit{Gaia} $P$, $e$, or $\omega$ also have $S<0.587$.

For the two SB2 HB systems with correct \textit{Gaia} periods and eccentricities we used the \texttt{PHysics Of Eclipsing BinariEs} (\texttt{PHOEBE}) package, a Python package for modeling EB systems \citep{Prsa2005, 2016Prsa, 2020Conroy}, to determine the stellar masses and radii. We use the period and orbital parameters from the HB models as the initial conditions. We use the \textit{Gaia} $K_1$ and $K_2$ velocity amplitudes to constrain the mass ratio

\begin{equation}
    q=\frac{K_{1}}{K_{2}},\ \text{with uncertainty}\ \sigma_{q}^2=q^2{\biggl(\frac{\sigma_{K_{1}}^2}{K_{1}^2}+\frac{\sigma_{K_{2}}^2}{K_{2}^2}\biggr)}.
	\label{eq:MassR}
\end{equation}and the projected semi-minor axis of the companion 

\begin{equation}
    a_{2}\sin i=\frac{K_{2}P}{2\pi\sqrt{1-e^2}},\ \text{with uncertainty}\ \sigma_{a_{2}sini}=a_{2}\sin i\frac{\sigma_{K_{2}}}{K_{2}}.
	\label{eq:SemiMinor}
\end{equation} Only the uncertainties in $K_1$ and $K_2$ are important. We initialize the \texttt{PHOEBE} MCMC walkers near the values from the initial conditions, use Gaussian priors on $q$ and $a_2 \sin i$, and then run the MCMC minimization with twenty walkers and five hundred iterations.

\begin{figure*}
	\includegraphics[width=\textwidth]{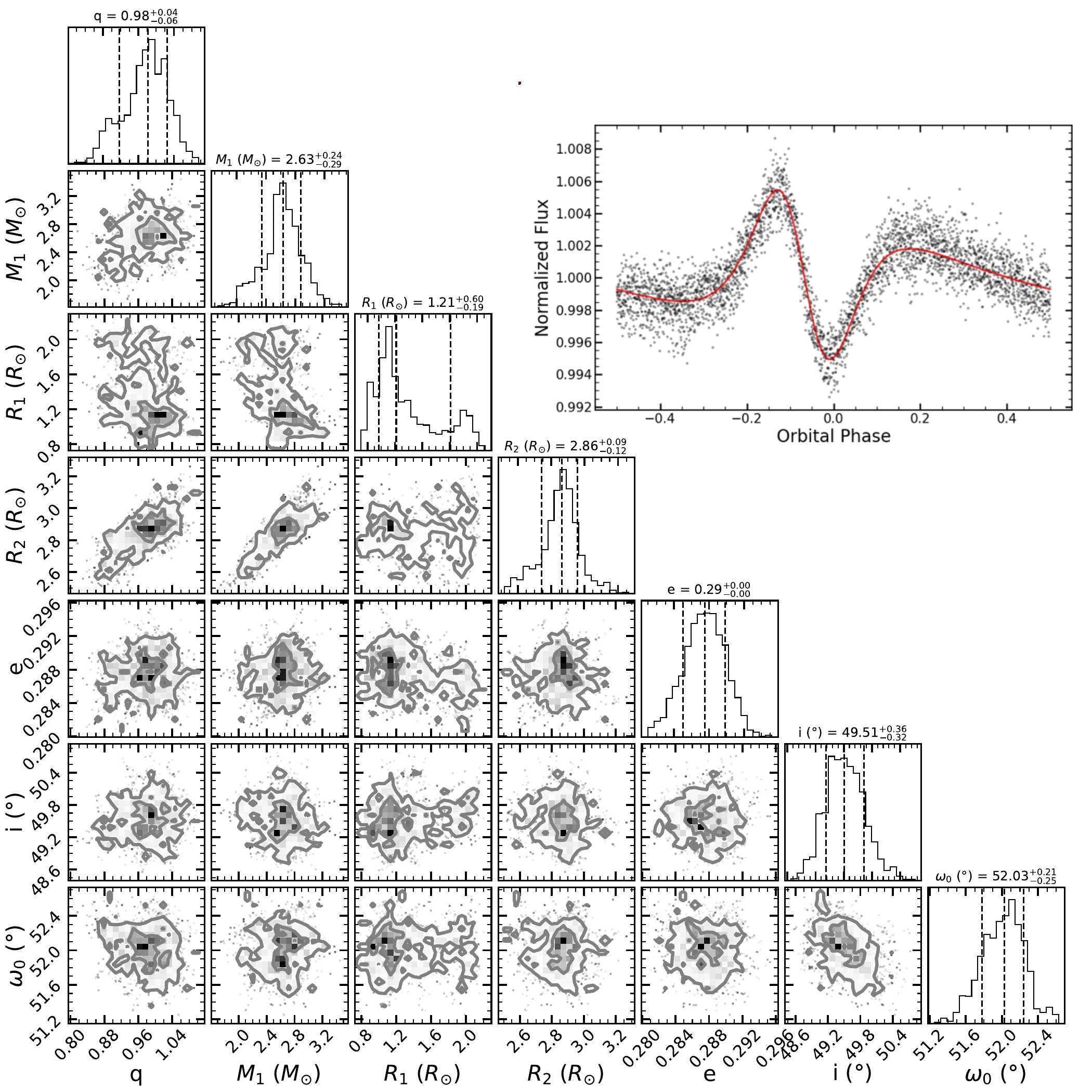}
    \caption{The posteriors from the \texttt{PHOEBE} MCMC models of TIC 98552498 for the mass ratio, q, primary mass, primary radius, secondary radius, eccentricity, inclination, and $\omega$. The projected probability distribution for each parameter and their 1$\sigma$ error bars (vertical lines) are shown at the top of each column. The inset shows the light curve and the best \texttt{PHOEBE} model.}
    \label{fig:Corner}
\end{figure*}

\begin{figure*}
	\includegraphics[width=\textwidth]{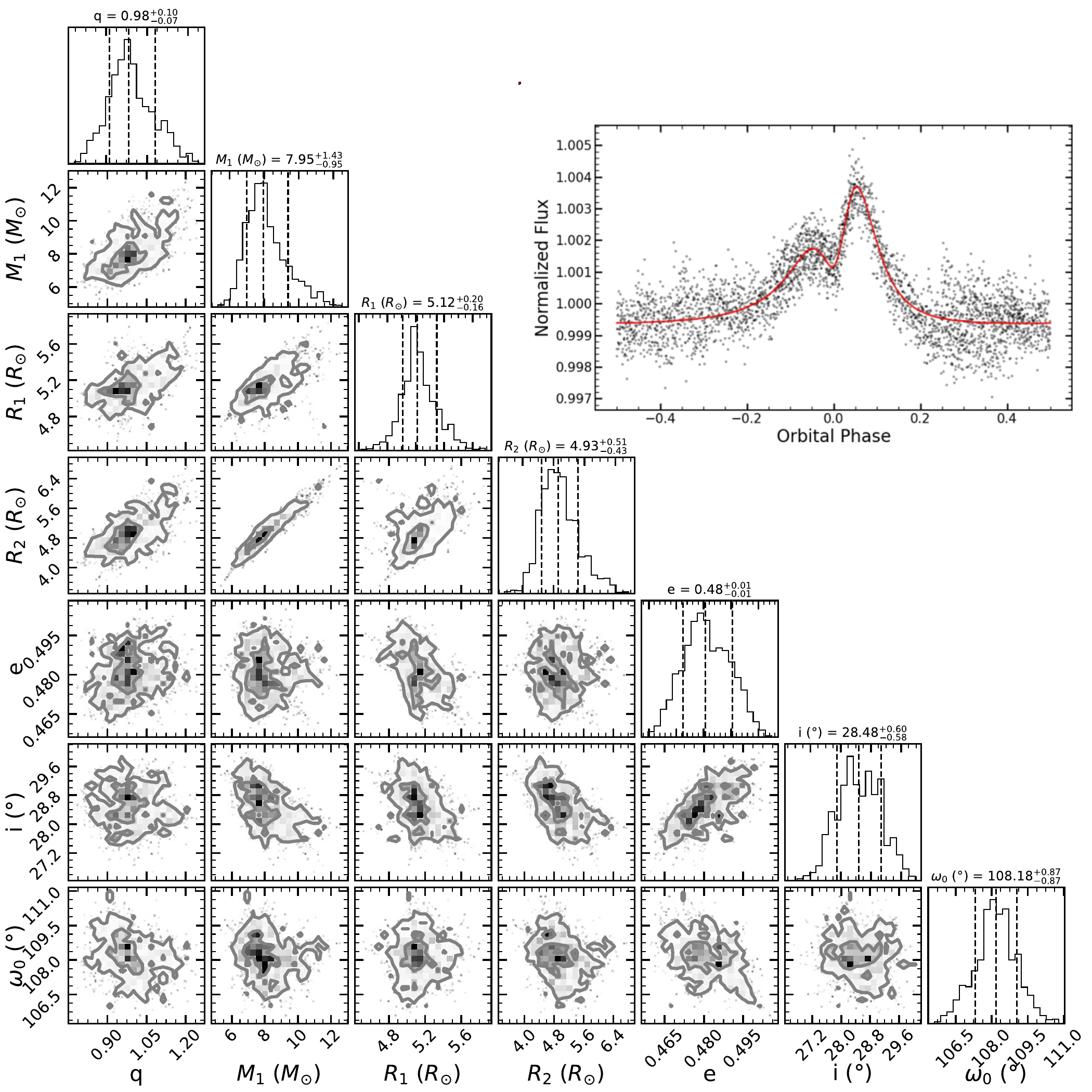}
    \caption{The posteriors from the \texttt{PHOEBE} MCMC models of TIC 76094846.}
    \label{fig:Corner2}
\end{figure*}

Fig.~\ref{fig:Corner} shows the posteriors for TIC 98552498. The mass and radius of the primary are (2.61 $\pm$ 0.27) $M_{\odot}$ and (1.41 $\pm$ 0.40) $R_{\odot}$, while the mass and radius of the secondary are (2.53 $\pm$ 0.29) $M_{\odot}$ and (2.85 $\pm$ 0.11) $R_{\odot}$. Fig.~\ref{fig:Corner2} shows the posteriors for TIC 76094846. The mass and radius of the primary are (8.19 $\pm$ 1.19) $M_{\odot}$ and (5.14 $\pm$ 0.18) $R_{\odot}$, while the mass and radius of the secondary are (8.11 $\pm$ 1.39) $M_{\odot}$ and (4.97 $\pm$ 0.47) $R_{\odot}$. Both systems are consistent with the masses and radii of massive stars measured from detached eclipsing binaries \citep{Torres2010}.

\begin{figure}
    \centering
    \begin{tabular}{c}
    \hspace{-2em}
    \includegraphics[width=\columnwidth]{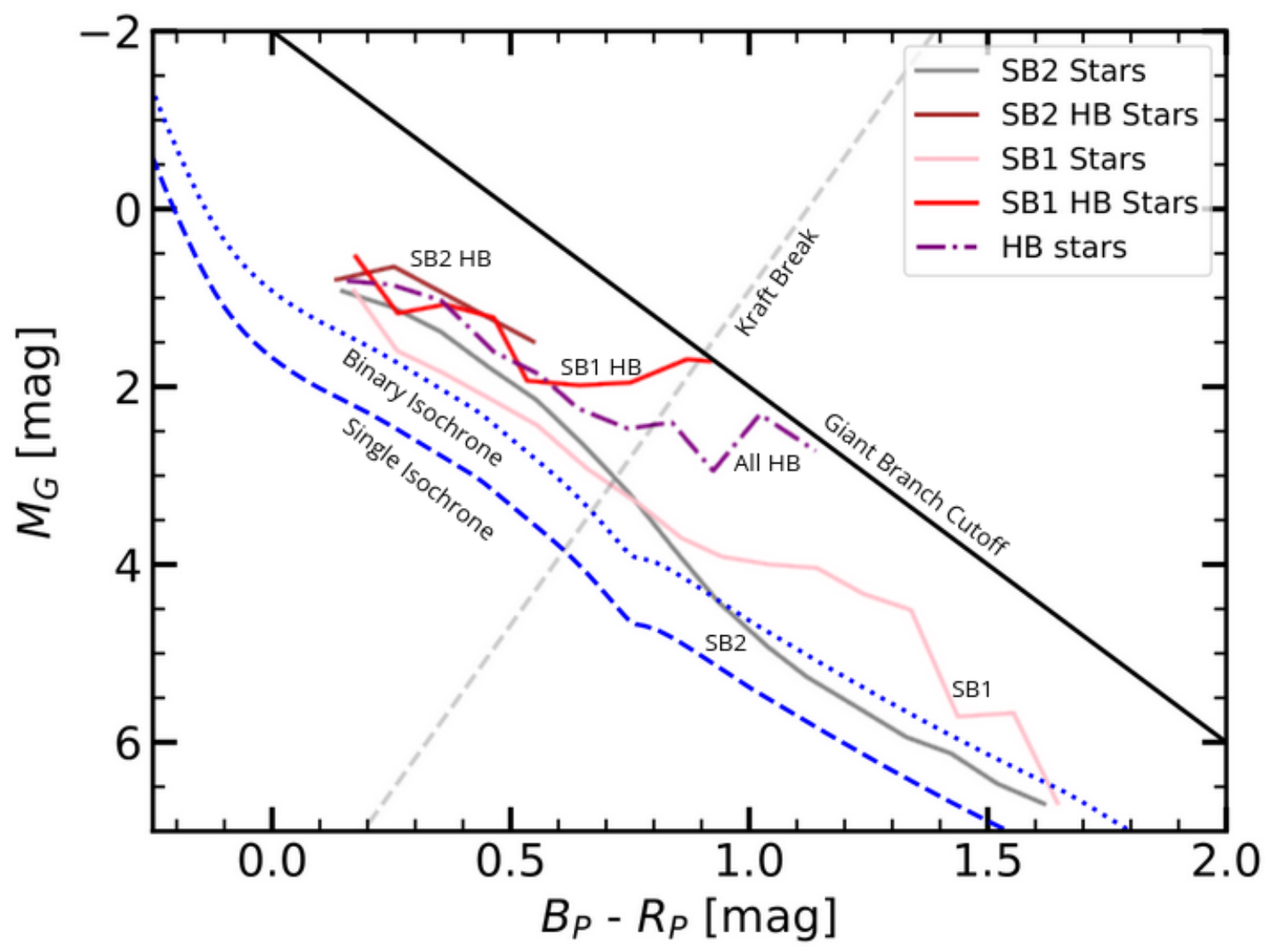} \\
    \end{tabular}
    \caption{Mean magnitudes in 0.1~mag wide colour bins for stars below the "Giant Branch Cutoff". The curves labeled SB1 and SB2 are for all the \textit{Gaia} binaries, the SB1 HB and SB2 HB curves are for the heartbeat stars found here, and the HB stars curve is for all HBs used in Fig.~\ref{fig:Mozeh} and Fig.~\ref{fig:CMD}. An 0.5 Gyr MIST isochrone \citep{2016Dotter, 2016Choi, 2011Paxton, 2013Paxton, 2015Paxton, 2018Paxton} is shown along with a binary "twin" isochrone a factor of two more luminous. The Kraft break line crosses the isochrone at $T_{eff} = 6550K$ \citep{1967Kraft, 1962Schatzman, 2024Beyer}}
    \label{fig:CMDB}
\end{figure}

\section{Discussion}
\label{sec:D}

We used the \textit{TESS} light curves of \textit{Gaia} spectroscopic binaries to identify one of the largest samples of HBs, finding 112 HBs among the 186,905 spectroscopic binaries in the \textit{Gaia} DR3 catalog, of these, 18 show eclipses, and 10 have TEOs. Out of the ten SB2 HBs, only two have orbital parameters in good agreement with the \textit{Gaia} values. The problem likely arises because of the sparse \textit{Gaia} RVs, the short periods, and that most of the systems are hot stars (Fig.~\ref{fig:CMD}). For the SB1 HBs, 85$\%$ of the orbits agreed with the \textit{Gaia} estimates. 

One of the original goals of this project was to find the SB2 HB systems and then model them to determine the component masses and radii. The problems with the SB2 solutions meant we could attempt this for only two systems. TIC 98552498 consists of a primary with mass (2.61 $\pm$ 0.27) $M_{\odot}$ and radius (1.41 $\pm$ 0.40) $R_{\odot}$, and a secondary with mass (2.53 $\pm$ 0.29) $M_{\odot}$ and radius (2.85 $\pm$ 0.11) $R_{\odot}$. TIC 76094846 consists of a primary with mass (8.19 $\pm$ 1.19) $M_{\odot}$ and radius (5.14 $\pm$ 0.18) $R_{\odot}$, and a secondary with mass (8.11 $\pm$ 1.39) $M_{\odot}$ and radius (4.97 $\pm$ 0.47) $R_{\odot}$. To analyze the remaining SB2 HBs, we require RV measurements from independent sources or the actual \textit{Gaia} RV measurements (which will be released in DR4).

There are two striking features of the distribution of HBs along the MS of the CMD shown in Fig.~\ref{fig:CMD}. First, the HBs seem to be more luminous at fixed color than their parent samples. Second, there seem to be many fewer HBs on the lower MS. To quantify these issues, we first applied a cut removing stars with $M_G < 4 (B_P-R_P-2)$ to eliminate most of the giant branch (Fig.~\ref{fig:CMDB}). This is imperfect for stars redder than roughly $B_P-R_P > 1$~mag, particularly for the SB1 systems. Fig.~\ref{fig:CMDB} also shows the 0.5~Gyr MIST isochrone (the "single" isochrone) and the same isochrone shifted to be 0.75~mag brighter (the "binary" isochrone), the shift expected for a binary of stellar twins. These are just to guide the eye since they are not a good match to the observed upper MS.

To examine the first point, we computed the median magnitudes of the samples in $B_P-R_P$ colour bins with a width of $0.1$~mag. The medians of the full (or period trimmed) SB1 and SB2 samples behave as expected on the upper MS. They roughly track the isochrones, and the SB2 magnitudes tend to be a little more luminous since having a more luminous companion will favor being able to obtain an SB2 solution. On the lower MS, the scatter of more luminous stars above the MS but below our cut against giants affects the SB1 sample more strongly than the SB2 sample because having an evolved primary makes measuring the velocity of the secondary less likely. The median absolute magnitudes of the HBs (SB1, SB2, or all HBs) are then more luminous than the SB2 sample. This strongly indicates that HBs are not just binaries, but binaries in which the primary has started to evolve. Following the logic in \citet{2025MacLeod}, this is a natural consequence of HB amplitudes being higher as stellar evolutionary time scales become shorter.

To examine the second point, we computed the fraction of HBs as a function of color. There are two reasons the HB fraction might drop as we examine stars lower on the MS. First, the redder, lower mass stars have longer and longer evolutionary time scales, which will lead to lower amplitudes \citep{2025MacLeod}. Second, stars below the Kraft break \citep[$T_K\simeq 6550$~K or $B_P-R_P \simeq 0.57$ for the isochrone in Fig.~\ref{fig:CMDB}, ][]{1967Kraft, 1962Schatzman, 2024Beyer} have convective envelopes and so should have shorter damping time scales. Both of these effects will lead to lower HB amplitudes. We have some ability to distinguish the two issues because the highest mass stars below the Kraft break \citep{1967Kraft, 1962Schatzman, 2024Beyer} can be old enough to have started to evolve off the MS. We computed the fractions of SB1 and SB2 HBs as a function of color again, but along lines parallel to the one labeled "Kraft break" in Fig.~\ref{fig:CMDB} in $B_P-R_P = 0.1$~mag bins. We report the median fractions and 90$\%$ confidence uncertainties or upper limits. 

\begin{figure}
    \centering
    \begin{tabular}{c}
	\includegraphics[width=\columnwidth]{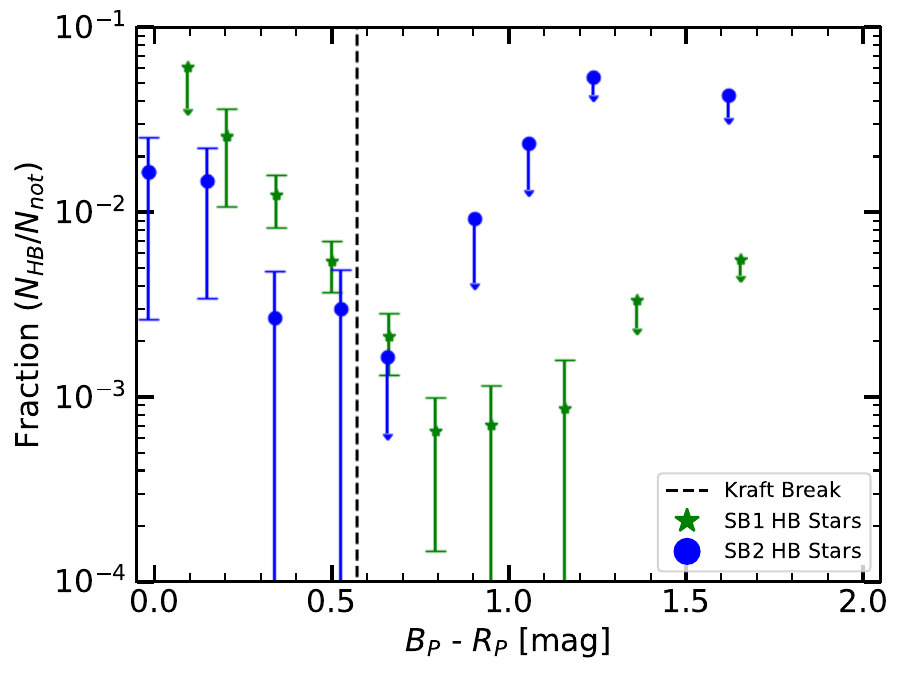} \\
    \includegraphics[width=\columnwidth]{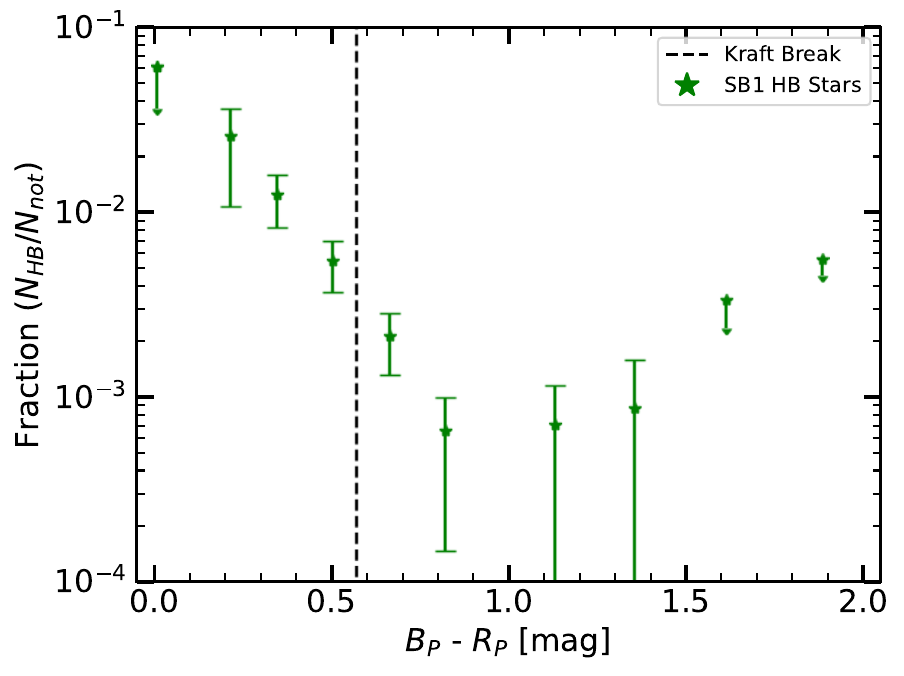} \\
    \end{tabular}
    \caption{The median fraction of HBs as a function of colour. Here the colour bins are done in slices parallel to the Kraft break \citep{1967Kraft, 1962Schatzman, 2024Beyer} line in Fig.~\ref{fig:CMDB}. The upper panel is for all SB2 systems and the SB1 systems with $P<13$ days. The lower panel shows the fraction for all SB1 systems with no period limit. The uncertainties and limits are at 90$\%$ confidence.}
    \label{fig:CMDF}
\end{figure}

The results shown in Fig.~\ref{fig:CMDF} are striking. The fraction of HBs drops very rapidly for redder, cooler stars starting above the Kraft break \citep{1967Kraft, 1962Schatzman, 2024Beyer} and then may be flattening for still cooler stars. The SB2 fractions and limits are in agreement with those for the SB1s. The absolute normalizations of the SB1 fractions change when we use the full SB1 sample for the normalization of the fraction, rather than just those with periods $<13$~days, but the trends are unchanged. On the upper MS, HBs are not rare - over 1$\%$ of \textit{Gaia} SB1 systems bluer than $B_P-R_P < 0.5$~mag are HBs.

Starting from a sample of binaries can simply be viewed as a means of increasing the efficiency of HB searches, particularly since many catalogs are dominated by shorter period systems. There are significant systematic effects in the catalogs which would need to be taken into account for any statistics as a function of period. This can be done more broadly by selecting \textit{Gaia} or APOGEE \citep[e.g., ][]{Abdurrouf2022} stars with significant velocity scatters \citep[e.g., ][]{Badenes2018} rather than requiring an orbit solution. It would be good to find a solution to the period search limitations of \textit{TESS}. Evolved HBs with longer periods may be detectable in the upcoming \textit{TESS} Cycle 8, which includes longer 54 day sectors and “rolled” sectors where some parts of the sky will be observed over consecutive sectors. It might also be possible to search for longer periods in \textit{TESS} and then use heavily binned All-Sky Automated Survey for Supernovae \citep[ASAS-SN, ][]{2014Shappee, 2017Kochanek} light curves to verify the periods and characterize the HB signal. While directly searching for such weak signals in ASAS-SN might well be problematic, \citet{2025Hon} have shown that binned ASAS-SN light curves can detect very weak signals.  

It is clear that there is considerable physics in characterizing and understanding the distribution of HBs as a function of orbital and stellar properties. What we qualitatively observe can largely be explained as the competition between evolutionary and dissipation rates, as discussed by \citet{2025MacLeod}. High mass stars evolve faster and have less dissipation, leading to larger numbers of HBs. There are, however, a large number of variables (mass, evolutionary state, orbit, etc.), which means that significantly larger samples will be needed to characterize the physical parameter space of HBs~fully.
\vspace{-1mm}

\section*{Acknowledgments}

JC thanks Tharindu Jayasinghe and the Time Domain Research Group for their guidance, advice, and support. JC also thanks Collin Christy for his assistance through the years. DMR acknowledges support from the OSU Presidential Fellowship. CSK, KZS, and DMR are supported by NSF grants AST-2307385 and 2407206. ASAS-SN is funded by Gordon and Betty Moore Foundation grants GBMF5490 and GBMF10501 and the Alfred P. Sloan Foundation grant G-2021-14192. Thanks to Dr. J. Tayar for a communication on an earlier version of this manuscript.

Our work presents results from the European Space Agency space mission Gaia. Gaia data are being processed by the Gaia Data Processing and Analysis Consortium (DPAC). Funding for the DPAC is provided by national institutions, in particular the institutions participating in the Gaia MultiLateral Agreement. This paper also includes data collected by the TESS mission. Funding for the TESS mission is provided by the NASA’s Science Mission Directorate. 

%The full version of Table~\ref{tab:table} is available in the online supplemental material.

\clearpage
\bibliographystyle{mnras}
\bibliography{hbs} % if your bibtex file is called example.bib

\end{document}